\documentclass[twocolumn,groupedaddress,amsmath,aps,longbibliography,nofootinbib]{revtex4-1}
\usepackage{hyperref}
\usepackage{graphicx}
\usepackage{amsmath,amsfonts}
\usepackage{dcolumn}
\usepackage{bm}
\usepackage{color}
\usepackage{multirow}
\usepackage{float}
\usepackage{url}
\usepackage{physics}
\usepackage{subfigure}
\usepackage[utf8x]{inputenc}

\begin{document}

\title{Pressure effects on the electronic structure, phonons, and superconductivity of noncentrosymmetric ThCoC$_2$}
\author{Gabriel Kuderowicz}
\author{Pawe\l{} W\'ojcik}
\author{Bartlomiej Wiendlocha}
\affiliation{Faculty of Physics and Applied Computer Science, AGH University of Science and Technology, Aleja Mickiewicza 30, 30-059 Krakow, Poland}
\email{wiendlocha@fis.agh.edu.pl}

\date{\today}

\begin{abstract}
The electronic structure, phonons, electron-phonon coupling, and superconductivity are theoretically studied in noncentrosymmetric superconductor ThCoC$_2$ as a function of pressure, 
in the pressure range 0 - 20 GPa.
We found that the electronic band splitting induced by the spin-orbit coupling is enhanced under pressure. 
In spite of the overall stiffening of the crystal lattice, the electron-phonon coupling constant $\lambda$ increases with pressure, from 0.583 at 0 GPa to 0.652 at 20 GPa. 
If the isotropic Eliashberg electron-phonon coupling theory is used to simulate the effect on the critical temperature $T_c$, such an increase in $\lambda$ results in a substantial increase of $T_c$ from 2.5~K at 0 GPa to 4~K at 20 GPa. {This shows that examining the effect of pressure offers a chance to resolve the pairing mechanism in ThCoC$_2$.}
\end{abstract}

\maketitle

\section{Introduction}
ThCoC$_2$ is a low-temperature noncentrosymmetric superconductor with the critical temperature $T_c \simeq 2.5$~K~\cite{Grant2014} and a puzzling nature of the superconducting state. The pairing mechanism has not been resolved yet, and the
thermodynamic properties of the superconducting phase strongly deviate from the predictions of the Bardeen-Cooper-Shrieffer theory (BCS)~\cite{BCS1957}. 
Among them, we may list:
(i) a non-exponential temperature dependence of the electronic specific heat $C_e(T)$, with the reduced specific heat jump of $\Delta C_e/\gamma T_c = 0.86$, much lower than the BCS value of 1.43 \cite{Grant2014};
(ii) the positive curvature of the temperature dependence of the upper critical magnetic field $H_{c2}(T)$;
(iii) a non-s-wave temperature dependence of the magnetic field penetration depth $\lambda_L(T)$ \cite{Bhattacharyya2019}. The measured $\lambda_L(T)$ was fitted \cite{Bhattacharyya2019} assuming a $d-$wave nodal superconducting gap. 
Moreover, spin fluctuations were suggested to be the pairing mechanism in ThCoC$_2$~\cite{Bhattacharyya2019}.

Very interesting results were also obtained when the material was doped with nickel~\cite{Grant2017}.
Critical temperature in ThCo$_{1-x}$Ni$_x$C$_2$ increased with $x$ reaching $T_c = 12$~K for $x=0.4$, with no signs of magnetism and a gradual suppressing of the non-BCS features of the superconducting phase towards a more conventional fully gapped superconductor~\cite{Grant2017}.

To shed more light on the nature of superconductivity in ThCoC$_2$, we have recently 
presented two theoretical works~\cite{Kuderowicz2021, Kuderowicz2021ni}.
The electronic structure, phonons and the electron-phonon coupling were computed in Ref.~\cite{Kuderowicz2021}
using the density functional perturbation theory~\cite{Giannozzi2009,QE-2017,dfpt}.
We found that strong spin-orbit coupling combined with the lack of inversion symmetry leads to a large band splitting with an average value of $\overline{\Delta E}_{SOC} \approx$ 150 meV, 
comparable to that observed in the triplet superconductors CePt$_3$Si and Li$_2$Pt$_3$B (see \cite{Smidman2017} and references therein). 
That indicates that the conventional $s$-wave gap symmetry will be strongly perturbed, as the pairing inside the spin-split bands in the strong spin-orbit coupling limit requires the odd parity of the gap with respect to the $\mathbf{k}\rightarrow -\mathbf{k}$~\cite{mixedstate2}.
As far as the pairing mechanism is concerned, the computed electron-phonon coupling constant $\lambda = 0.59$  was large enough to propose the electron-phonon interaction to be responsible for superconductivity in ThCoC$_2$~\cite{Kuderowicz2021}.
However, our calculations within the isotropic Eliashberg formalism \cite{Eliashberg1960} did not explain the experimental non-BCS features of ThCoC$_2$, supporting the non-$s$-wave picture of its superconductivity, but with the possibility of the electron-phonon coupling mechanism~\cite{Kuderowicz2021}.

Electron-phonon coupling hypothesis was further strengthened by calculations of the effect of Ni doping presented in Ref.~\cite{Kuderowicz2021ni}.
Using the simplified rigid muffin-tin approximation (RMTA), the evolution of the electron-phonon coupling constant in ThCo$_{1-x}$Ni$_x$C$_2$ was analyzed.
Although due to the usage of the RMTA approach, the calculated values of $\lambda(x)$ were systematically underestimated when compared to the experimental estimates, 
the strong increase of $\lambda(x)$ was found, which remained in a qualitative agreement with the experimental observation of the increase in $T_c(x)$ under the assumption of the electron-phonon coupling mechanism of superconductivity in ThCo$_{1-x}$Ni$_x$C$_2$.

In this work, we study the effects of pressure on the electronic structure, phonons, electron-phonon interaction, and superconductivity in ThCoC$_2$, 
which should help to determine the pairing interaction if the experimental studies are undertaken.
In this context, it is important to recall the interesting pressure dependence of superconductivity in the sister isostructural superconductor $\mathrm{LaNiC_2}$. 
The external pressure initially increases the critical temperature from 2.8 K at 0 GPa to 3.8 at 4 GPa and later decreases $T_c$ to suppress superconductivity above 7 GPa~\cite{Katano2014}.
The calculated electron-phonon coupling parameter $\lambda$ and critical temperature $T_c$, on the other hand, were found to monotonously increase with pressure~\cite{lanic2}, in agreement with the initial increase in $T_c(p)$ 
but showing the non-phononic origin of the suppression of superconductivity at larger pressures. 

\section{Computational details}\label{sec:comp_details}

Calculations were done for p = 5, 10, 15, and 20 GPa, and to be consistent with our earlier work for ambient pressure, all computational details were kept unchanged. 
Some of the results for p = 0 GPa are recalled here for a convenient comparison.
The {\sc Quantum Espresso} package~\cite{Giannozzi2009,QE-2017} was used, 
with the Perdew-Burke-Ernzerhof (PBE) generalized gradient approximation for the exchange-correlation functional~\cite{pbe}
and the Rappe-Rabe-Kaxiras-Joannopoulos (RRKJ) ultrasoft pseudopotentials~\cite{ThCoC2pseudos,dalcorso}. 
Electronic structure was calculated in the scalar-relativistic and fully relativistic way (i.e., including the spin-orbit coupling). 
Phonons and electron-phonon coupling functions were calculated in the scalar-relativistic way, as the inclusion of SOC had no important effect on the phonon structure 
nor on the electron-phonon interaction~\cite{Kuderowicz2021}.

\begin{table}[b]
\caption{Evolution of cell parameters with applied pressure. Atomic positions (in crystal coordinates) are in a form: Th (0,0,u), Co (0.5,0,v), C (0.5,$\pm$y,z).\label{tab_cell_px}}
\centering
\begin{ruledtabular}
\begin{tabular}{cccccccc}
	p (GPa) & a (\AA) & b (\AA) & c (\AA) & u & v & y & z\\
	\hline
	0 & 3.8214 & 4.5376 & 6.0708 & -0.0014 & 0.6041 & 0.1561 & 0.3007\\
	5 & 3.7781 & 4.5120 & 6.0219 & -0.0011 & 0.6037 & 0.1569 & 0.3007\\
	10 & 3.7392 & 4.4885 & 5.9786 & -0.0009 & 0.6034 & 0.1575 & 0.3007\\
	15 & 3.7031 & 4.4670 & 5.9408 & -0.0006 & 0.6032 & 0.1581 & 0.3007\\
	20 & 3.6705 & 4.4467 & 5.9063 & -0.0004 & 0.6030 & 0.1586 & 0.3007\\
	\end{tabular}
\end{ruledtabular}
\end{table}

\begin{figure}[t]
	\centering
	\includegraphics[width=0.99\columnwidth]{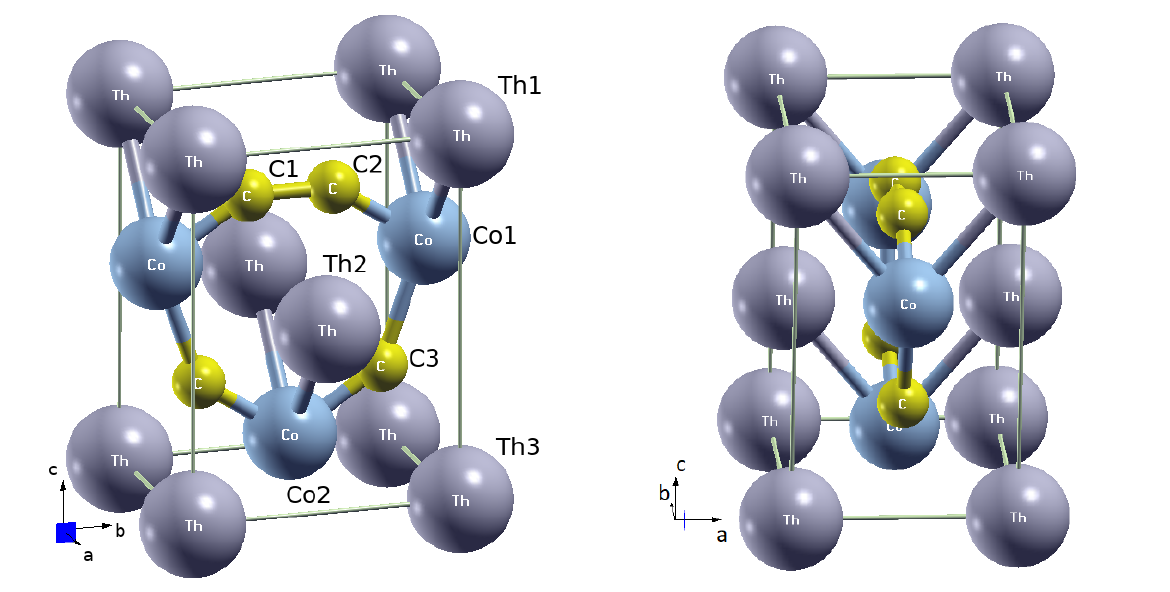}
	\caption{The noncentrosymmetric unit cell of $\mathrm{ThCoC_2}$ visualized with {\sc xcrysden}~\cite{Kokalj2003}. Numbers are used in Table~\ref{tab_atom_dist} to define interatomic distances. In the right panel we may distinguish 
	the layers of Th separated by the layers of Co-C rings.	\label{plot_struct}}	
\end{figure}

Crystal unit cell parameters and atomic positions were theoretically optimized, starting from the p = 0 GPa experimental values~\cite{Gerss1986}.
Electronic structure was then calculated using a Monkhorst-Pack grid of $12^3$ {\it k}-points, whereas the Fermi surface was calculated on a $18^3$ mesh. 
Wavefunction energy and charge density cutoffs were set to $130$~Ry and $1300$~Ry, respectively. 
Dynamical matrices were calculated on a 4x4x4 {\it q-}point mesh and the interatomic force constants (IFC) and phonon dispersion relations were computed using the Fourier interpolation technique. 
Finally, the Eliashberg electron-phonon interaction functions $\alpha ^2 F(\omega)$ were computed using the self-consistent first order variation of the crystal potential~\cite{dfpt}.
The spectral functions $\alpha ^2 F(\omega)$ were then used to 
determine the pressure dependence of the electron-phonon coupling (EPC) parameter $\lambda$.
The thermodynamic properties of the superconducting phase were determined using the isotropic Eliashberg formalism to see how they evolve under pressure, but one has to be aware that the superconducting state in ThCoC$_2$ at p = 0 GPa could not be accurately described within this formalism. The isotropic theory is used here to simulate the trend in $T_c(p)$, as well as to prepare the grounds for a discussion of future experimental results.

\begin{figure}[t]
	\centering
	\includegraphics[width=0.99\columnwidth]{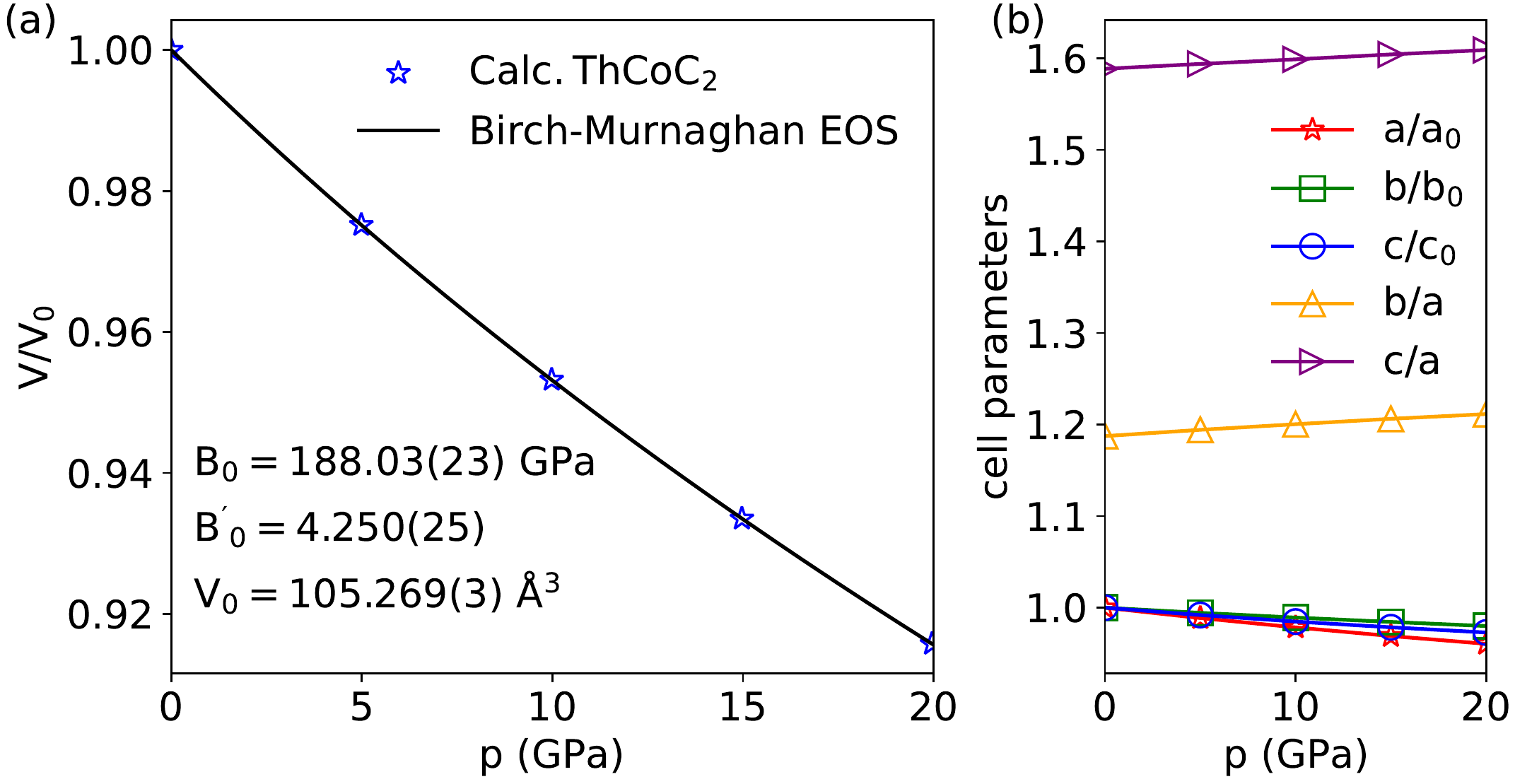}
	\caption{(a) Birch-Murnaghan equation of state fitted to $\mathrm{ThCoC_2}$ unit cell volumes under pressure. (b) Evolution of cell parameters (lines are guides for the eye). \label{plot_eos}}	
\end{figure}

\begin{table}[b]
\caption{Distances between neighboring atoms (\AA) in $\mathrm{ThCoC_2}$ under pressure p (GPa). Atoms are numbered as shown in Fig. \ref{plot_struct}. $\mathrm{\Delta d=d_{20\ GPa}-d_{0\  GPa}}$.\label{tab_atom_dist}}
\centering
\begin{ruledtabular}
\begin{tabular}{cccccccc}
	\multirow{2}{*}{p} & C1 - & C2 - & C2 - & Co1 - & Co1 - & Th1 - & Th2 -\\
	 & C2 & Th1 & Th2 & C2 & C3 & Co1 & Co1\\
	\hline
    0  & 1.4165 & 2.7442 & 2.7412 & 1.9646 & 1.9734 & 3.0641 & 3.0345\\
    5  & 1.4156 & 2.7184 & 2.7152 & 1.9504 & 1.9571 & 3.0384 & 3.0094\\
    10 & 1.4142 & 2.6954 & 2.6920 & 1.9377 & 1.9428 & 3.0154 & 2.9868\\
    15 & 1.4128 & 2.6746 & 2.6705 & 1.9260 & 1.9306 & 2.9949 & 2.9659\\
    20 & 1.4108 & 2.6555 & 2.6513 & 1.9153 & 1.9194 & 2.9759 & 2.9469\\
    \hline
    $\mathrm{\Delta d}$ & -0.0057 & -0.0887 & -0.0899 & -0.0493 & -0.0540 & -0.0882 & -0.0876\\
    $\mathrm{\frac{\Delta d}{GPa}}$ & -0.0003 & -0.0044 & -0.0045 & -0.0025 & -0.0027 & -0.0044 & -0.0044\\
	\end{tabular}
\end{ruledtabular}
\end{table}

\section{Results and discussion}

\begin{figure*}[t]
	\centering
	\includegraphics[width=0.85\textwidth]{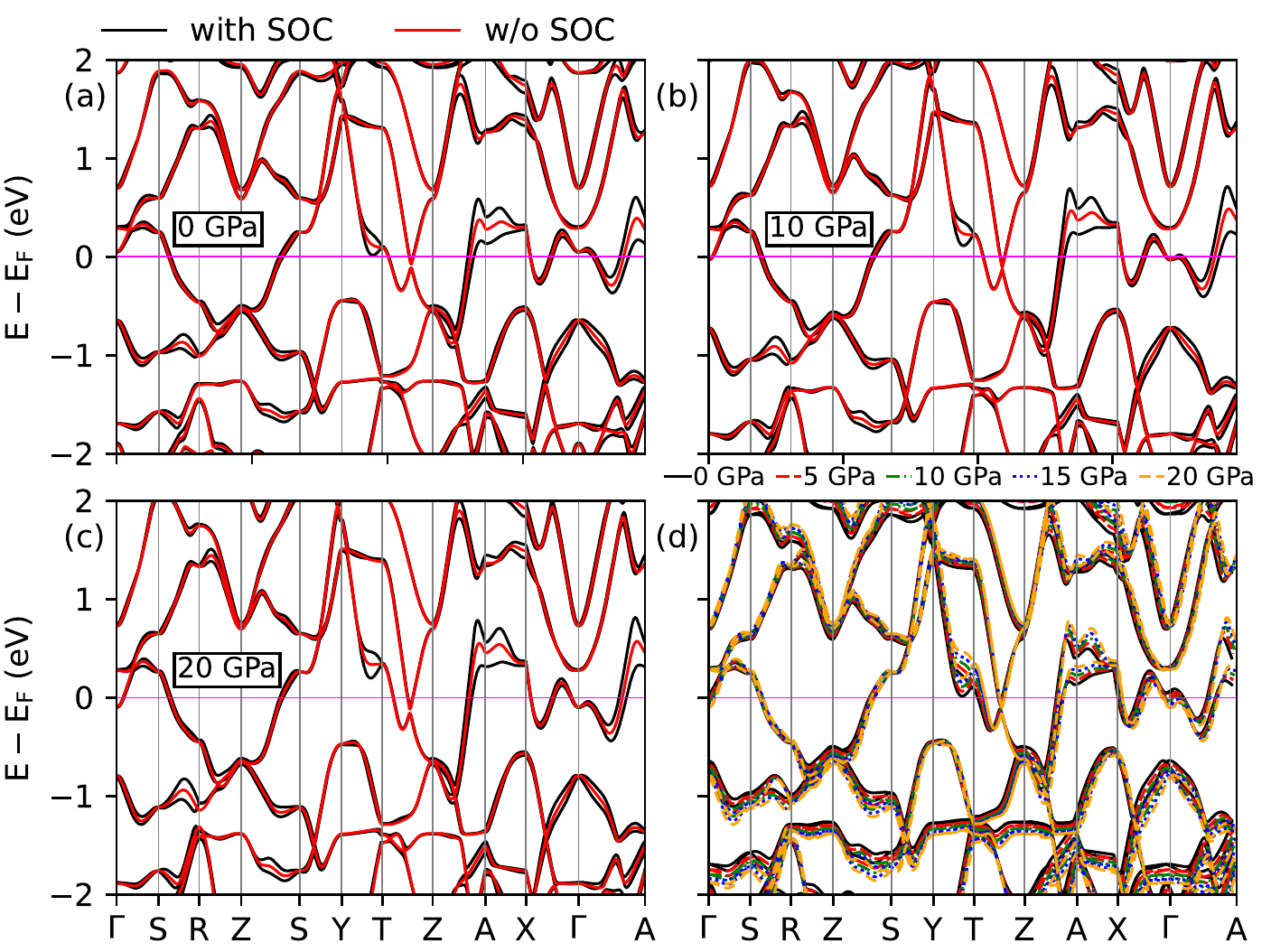}
	\caption{Electronic dispersion relation of $\mathrm{ThCoC_2}$ under selected pressures.
	{In panels (a-c) the scalar-relativistic and fully relativistic results are plotted for 0, 10 and 20 GPa, while in (d) the relativistic results for all pressures are gathered to visualize their relative differences.}\label{plot_elbands}}
\end{figure*}

\subsection{Crystal structure}
ThCoC$_2$ crystallizes in the base-centered orthorhombic structure \textit{Amm2}, spacegroup no. 38, shown in Fig.~\ref{plot_struct}.
In the structure we may distinguish the layers of Th separated by the layers of Co-C rings, stacked along $a$ axis.
Evolution of the unit cell parameters is presented in Table~\ref{tab_cell_px} and Fig.~\ref{plot_eos}.
The strongest compression of the unit cell under pressure is seen in the $a$ direction, i.e. perpendicular to the aforementioned Th and Co-C layers, whereas
the most resistant to pressure is $b$ direction, along the covalent Co-C and C-C bonds. As a consequence, $b/a$ and $c/a$ ratios increase with pressure.
Atomic positions change only slightly, and to illustrate how the crystal structure evolved, the distances between neighboring atoms are shown in Table \ref{tab_atom_dist}. 
The carbon-carbon bond is the stiffest as its 
bond length shortens an order of magnitude less than in the other pairs. 
Distances between cobalt and carbon decrease roughly two times less than for the carbon - thorium and thorium - cobalt pairs.

The change in the unit cell volume with pressure is shown in Fig.~\ref{plot_eos}(a) where it is fitted using the Birch-Murnaghan equation of state \cite{Birch1947}. The computed bulk modulus is equal to $B = 188$ GPa.

\begin{figure*}[t]
	\centering
    \includegraphics[width=0.99\textwidth]{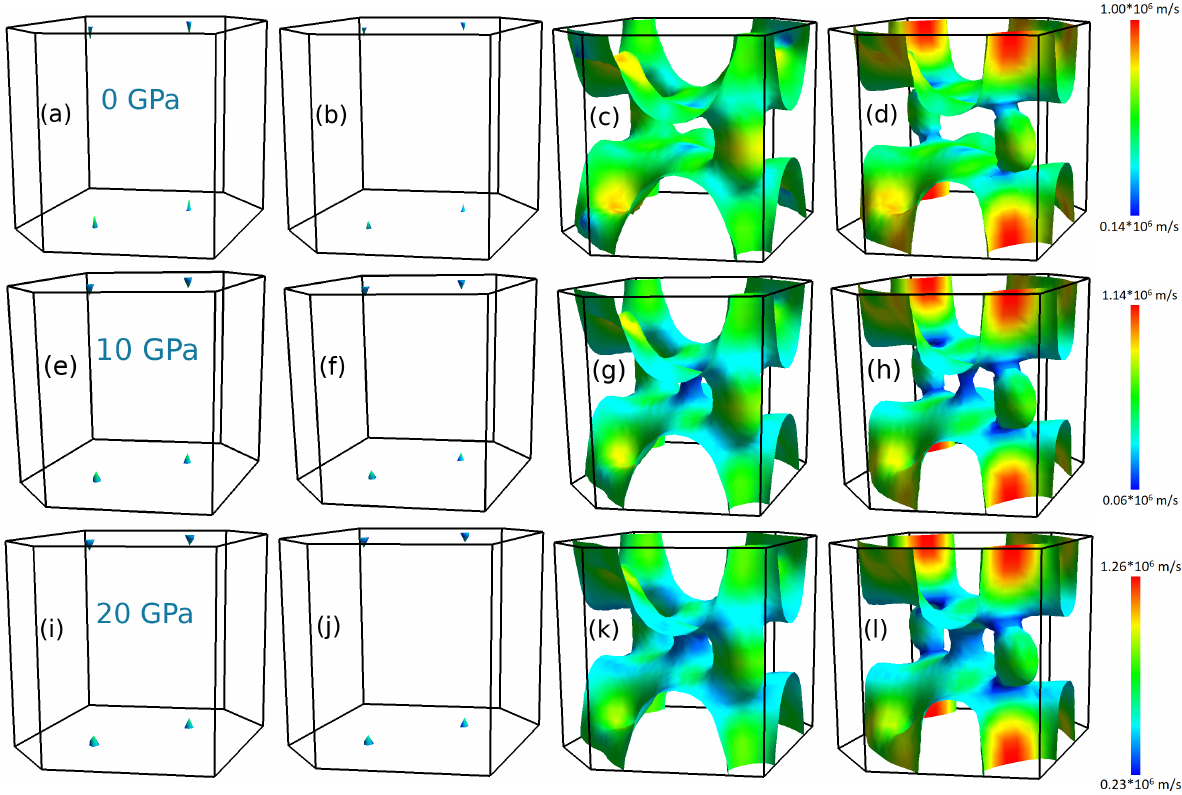}
	\caption{The Fermi surface (FS) of $\mathrm{ThCoC_2}$ calculated with SOC. Top row shows FS sheets for 0 GPa (a-d), middle for 10 GPa (e-h) and bottom for 20 GPa (i-l). FS visualized using {\sc fermisurfer}~\cite{fermisurfer}. Color marks the Fermi velocity.\label{plot_fs}}
\end{figure*}

\subsection{Electronic structure}\label{sec:electronic_structure}

Evolution of the electronic dispersion relations in ThCoC$_2$ with pressure is presented in Fig.~\ref{plot_elbands}. 
{Panels (a-c) show dispersion relations for 0, 10 and 20 GPa, while panel (d) gathers all results in one graph to visualize their relative differences. In a similar way we will present changes in electronic densities of states, phonon dispersions, phonon densities of states and Eliashberg functions.}
As discussed in~\cite{Kuderowicz2021}, at ambient pressure in the scalar-relativistic case, two bands cross the Fermi level forming two sheets of the Fermi surface (FS), with one large, dominating FS sheet and the other consisting of two small electron pockets. 
When the spin-orbit coupling is taken into account, due to the lack of the inversion center, the band structure is split, forming four FS sheets, shown for p = 0, 10, and 20 GPa in Fig.~\ref{plot_fs}.
The largest change in $E({\bf k})$ relation due to pressure is observed near $\mathrm{\Gamma}$ and T points. 
The electron band at $\mathrm{\Gamma}$, which is just above $E_F$ for  p = 0 GPa, is pushed down and eventually crosses the Fermi level, contributing to FS by forming an additional tube along $k_z$ direction, seen in Fig.~\ref{plot_fs} for p = 10 GPa.

\begin{figure}[b]
	\centering
	\includegraphics[width=0.99\columnwidth]{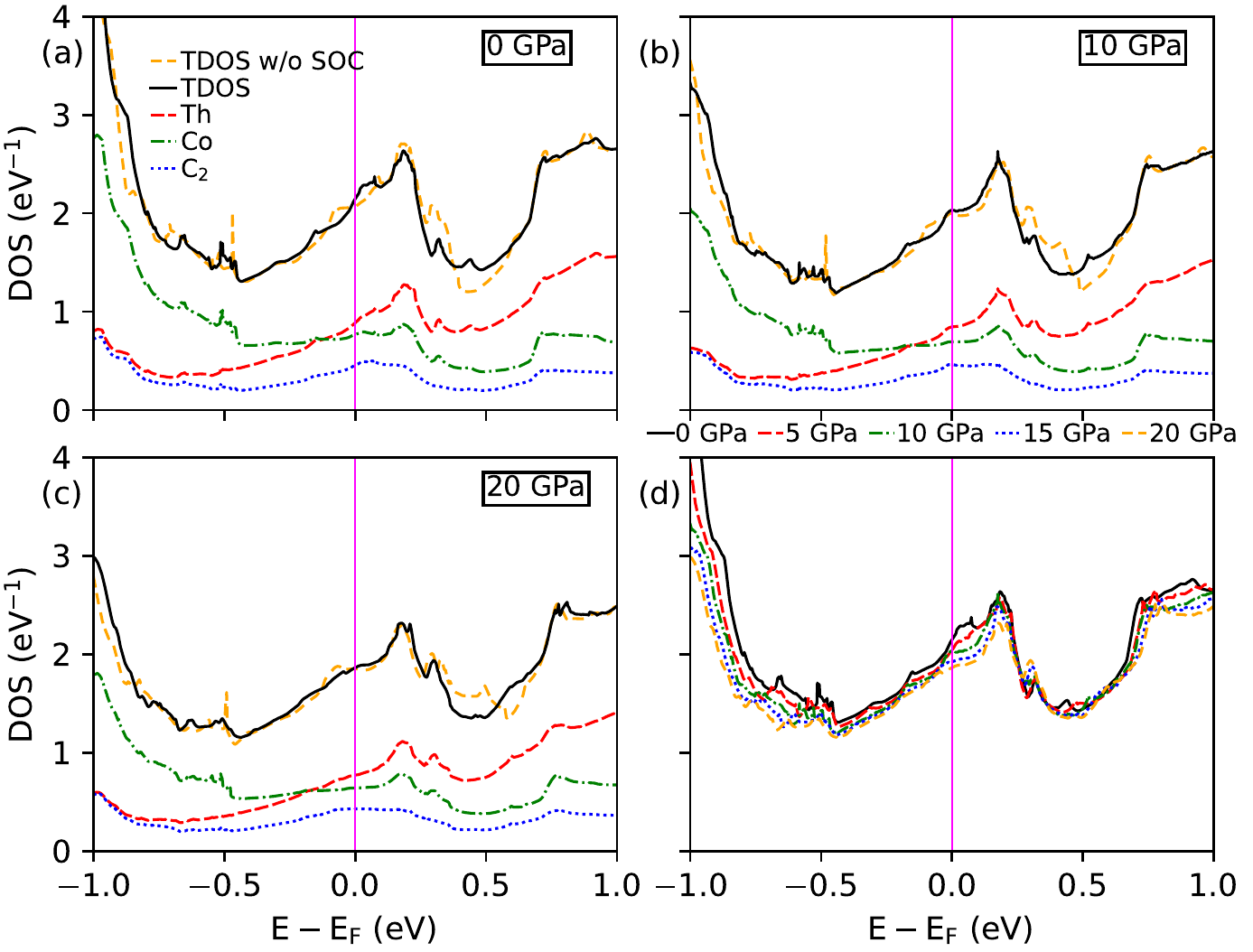}
	\caption{Total and partial electronic density of states of $\mathrm{ThCoC_2}$ under selected pressures. Magenta line marks the Fermi energy. {In panels (a-c) fully relativistic total and partial DOS results are plotted for 0, 10 and 20 GPa and compared to the scalar-relativistic total DOS, while in (d) the relativistic results for all pressures are gathered to visualize their relative differences.}
	\label{plot_eldos}}
\end{figure}

{Orbital characters of the main Fermi surface sheet for pressures of 0, 10  and 20 GPa are shown in Supplemental Material~\cite{supplemental}. Contribution to Fermi surface is mostly from Th-6d, Co-3d and C-2p orbitals, with some contribution from Th-f states, in agreement with contributions to total DOS at $E_F$ discussed below (see Fig.~\ref{plot_eldos}). The anisotropy of orbital character of FS is moderate and is enhanced under pressure -- the tube which is formed in the central part of FS above 10 GPa has mostly C-p and Th-f character, which were more uniformly distributed at ambient pressure. That may enhance the anisotropic properties of the superconducting phase under pressure in ThCoC$_2$.}

\begin{figure}[t]
	\centering
	\includegraphics[width=0.99\columnwidth]{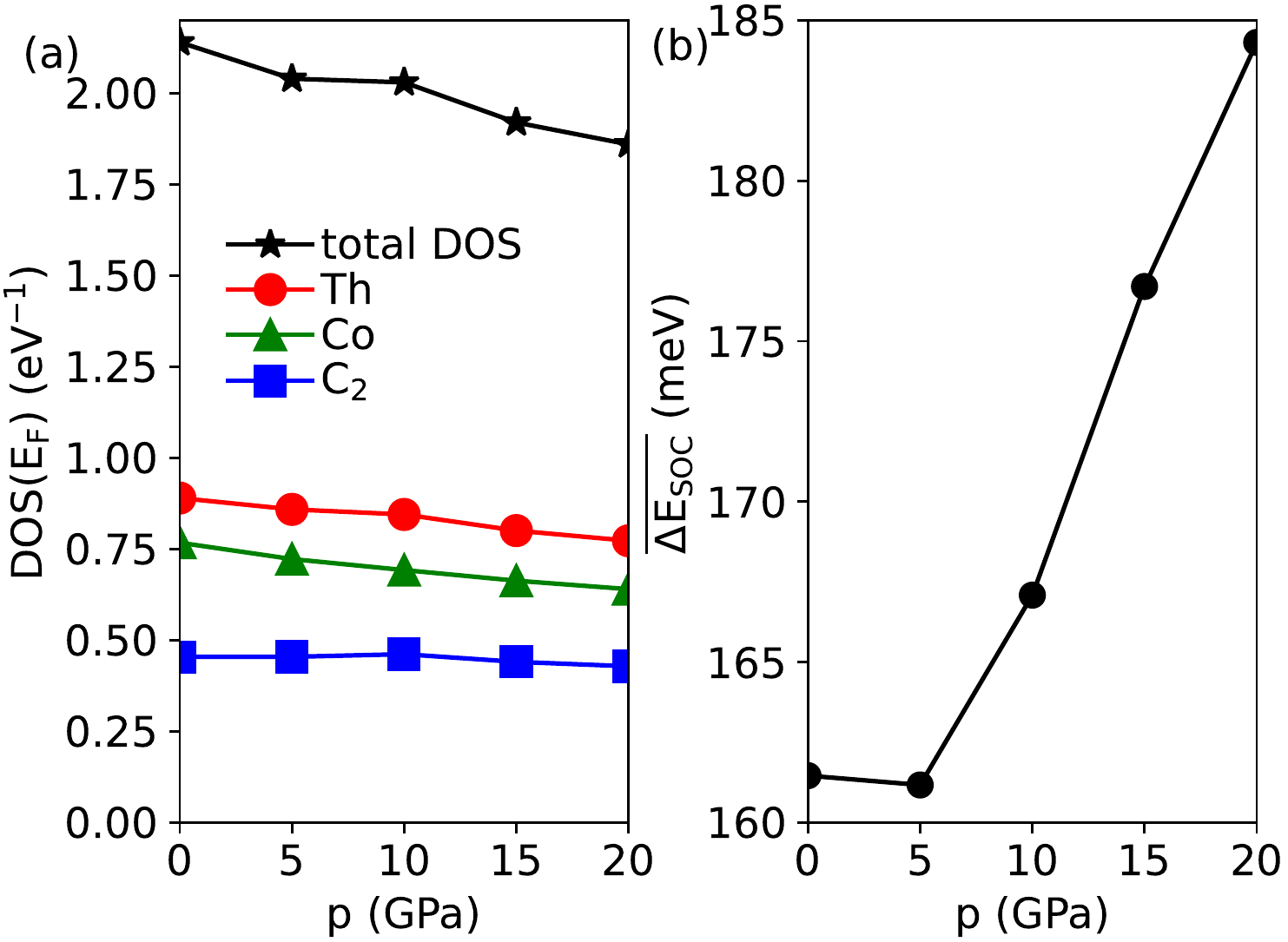}
	\caption{(a) Total and partial electronic density of states at the Fermi energy of $\mathrm{ThCoC_2}$ under selected pressures. (b) Average SOC splitting of the large Fermi surface sheet. Lines are guides for the eye.\label{plot_ef}}
\end{figure}

The average value of $E({\bf k})$ splitting due to SOC for the states located around the Fermi level as a function of pressure is shown in Fig.~\ref{plot_ef}(b).  
The $\overline{\Delta E}_{SOC}$ starts to increase above 5 GPa when the additional, $\Gamma$-centered tube starts to contribute to FS. 
The growth is approximately linear with the a rate of 1.5 meV/GPa, reaching 185 meV at 20 GPa\footnote{The average splitting at p = 0 GPa reported here is 162 meV is slightly larger than the (rounded down) value of 150 meV reported in Ref.\cite{Kuderowicz2021} due to a denser $k$-mesh used to calculate the average.}.
That suggests that the unconventional features of superconductivity in ThCoC$_2$, likely driven by the strong antisymmetric spin-orbit coupling, should be even more pronounced under elevated pressure.

The electronic density of states (DOS) near the Fermi level is presented in Fig.~\ref{plot_eldos} and the change of DOS($E_F$) with pressure in Fig.~\ref{plot_ef}(a) and Table~\ref{tab_dosef}.
Under the pressure, the DOS curve around $E_F$ is flattened, due to pressure-enhanced hybridization we can observe lowering of maxima and rising of minima of DOS($E$).
The DOS($E_F$) gradually decreases with pressure with a ratio of $-1.1\times 10^{-2}$ eV$^{-1}/$GPa
which compares to $-2.7\times 10^{-3}$ eV$^{-1}/$GPa in LaNiC$_2$~\cite{lanic2},
thus, the effect is much stronger in ThCoC$_2$.
{The largest contributions to the total DOS at the Fermi level 
are from the 6d Th, 3d Co, and 2p C states, with a noticeable contribution from 5f states of Th. The major contributors to DOS$(E_F)$ do not change with pressure, however the relative importance of Th-5f states is growing with increasing pressure, due to the above-mentioned increase in the Fermi surface area with the Th-5f orbital character.} 

\begin{table}[htb]
\caption{Density of states of $\mathrm{ThCoC_2}$ at the Fermi energy.\label{tab_dosef}}
\centering
\begin{ruledtabular}
\begin{tabular}{ccccccc}
	$N(E_F)$ & p & \multirow{2}{*}{0} & \multirow{2}{*}{5} & \multirow{2}{*}{10} & \multirow{2}{*}{15} & \multirow{2}{*}{20}\\
	$\mathrm{(eV^{-1})}$ & (GPa) & & & & & \\
	\hline
	w/o SOC & total & 2.07 & 2.02 & 2.00 & 1.91 & 1.85\\
	with SOC & total & 2.14 & 2.04 & 2.03 & 1.92 & 1.86\\
	 & Th-6d & 0.59 & 0.55 & 0.53 & 0.50 & 0.48 \\
	 & Th-5f & 0.25 & 0.26 & 0.27 & 0.26 & 0.25\\
	 & Co-3d & 0.76 & 0.72 & 0.69 & 0.66 & 0.64\\
	 & C$_2$-2p & 0.44 & 0.44 & 0.45 & 0.43 & 0.42\\
	 
	\end{tabular}
\end{ruledtabular}
\end{table}

\subsection{\label{sec:eph}Phonons and electron-phonon coupling}

As we found for the zero pressure case~\cite{Kuderowicz2021}, spin-orbit coupling makes no significant changes for the phonon spectrum of ThCoC$_2$, as well as for the electron-phonon interactions. The computed scalar-relativistic value of the EPC constant for p = 0,  $\lambda = 0.583$, is only slightly smaller than the relativistic $\lambda = 0.590$. 
Because of this, we restricted our calculations of the pressure evolution of the dynamic properties and EPC in ThCoC$_2$ to the scalar-relativistic case. 
To make sure that SOC may be also neglected for phonons under pressure, giving the correct evolution of $\lambda(p)$,
calculations including SOC were done for a single case of p = 10 GPa and also here no important differences were found (see Supplemental Material~\cite{supplemental} for further details).
Similarly to the p = 0 case, at 10 GPa the scalar-relativistic $\lambda = 0.609$ is slightly smaller than the relativistic $\lambda = 0.618$. 
The relative effect of pressure on the EPC parameter between 10 GPa and 0 GPa, $\lambda(10)/\lambda(0)$, is 4.7\% increase, when computed with SOC, and 4.5\% increase, when SOC is neglected, thus the scalar-relativistic approximation is sufficient to drive conclusions for the evolution of $\lambda$ with pressure.

\begin{figure*}[htb]
	\centering
	\includegraphics[width=0.99\textwidth]{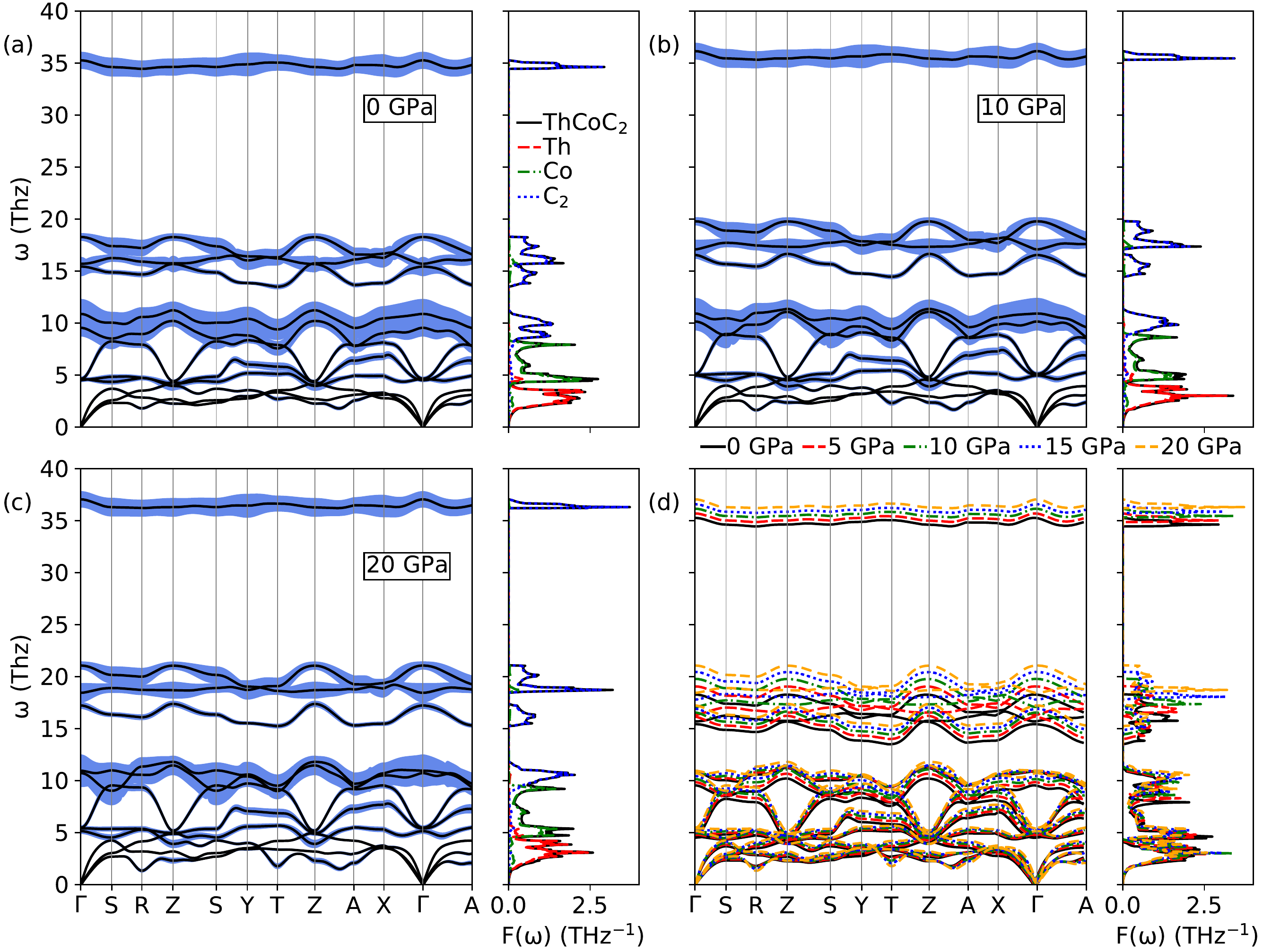}
	\caption{Phonon dispersion relations and phonon densities of states of $\mathrm{ThCoC_2}$ under selected pressures. {In panels (a-c) results are plotted for 0, 10 and 20 GPa and in the phonon dispersion plots blue shading shows the phonon linewidths (multiplied by 15). In panel (d) results for all pressures are gathered to visualize their relative evolution with pressure, here the phonon linewidths are not shown.}
	\label{plot_phbands}}
\end{figure*}

Figure~\ref{plot_phbands} presents the phonon dispersion curves 
$\omega(\mathbf{q})$ and the phonon density of states $F(\omega)$ for selected pressures. 
The characteristic structure of the phonon spectrum, driven by the large difference in atomic masses between Th, Co, and C, is the separation into three regions, seen in the partial $F(\omega)$ plots. The lowest-frequency part has the majority of Th vibrations, next Co phonon branches dominate and the high 
frequency part is contributed by carbon atoms. The highest single mode near $35$~THz (p = 0 GPa) is contributed by the C-C bond-stretching mode. 

\begin{figure}[b]
	\centering
	\includegraphics[width=0.99\columnwidth]{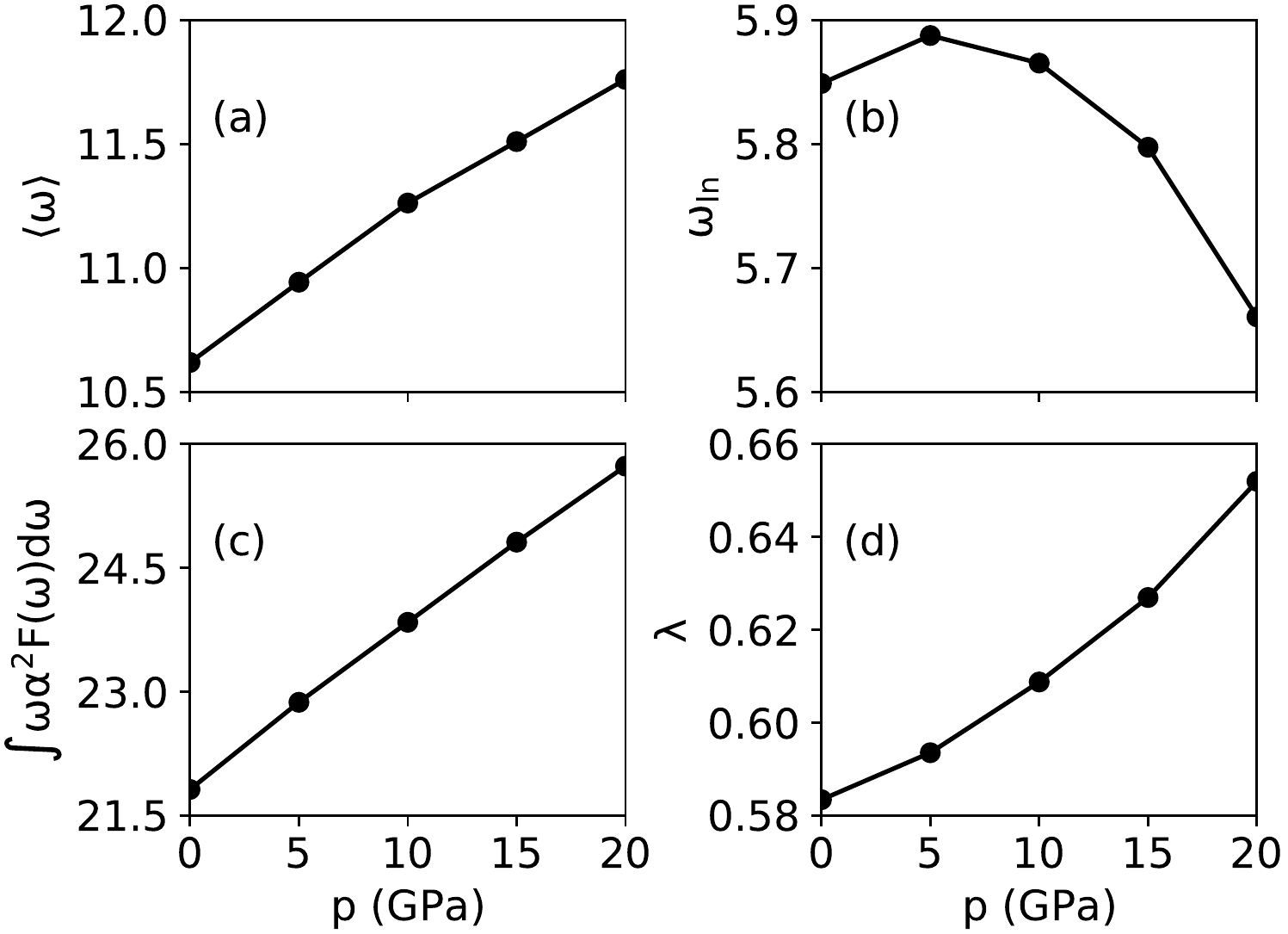}
	\caption{(a) Average phonon frequency, (b) logarithmic average frequency $\omega_{\rm ln}$ [eq.(\ref{eq:wln}, (c) integral $I$ [eq.(\ref{eq:I} and (d) electron-phonon coupling constant $\lambda$ of $\mathrm{ThCoC_2}$. \label{plot_meanfreqs}}
\end{figure}

Under elevated pressure, most of the phonon branches move towards higher frequencies, which is especially visible for carbon modes. 
The average phonon frequency increases with an approximate rate of 0.057 THz/GPa, reaching 11.76 THz at 20 GPa, as shown in Fig.~\ref{plot_meanfreqs}(a).
On the other hand, near R, T, and Z points, the lowest acoustic mode is softened. Similar softening  was found in LaNiC$_2$~\cite{lanic2}.
The real-space atomic displacements for these phonon modes are visualized in the Supplemental Material~\cite{supplemental}. In T point, where the effect is the strongest, Th vibrates along the $c$ axis, whereas Co and C move out of phase, along $a$.

Before moving to the discussion of the electron-phonon coupling, it is worth to recall how the quantities important for the interaction strength depend on the phonon frequencies. 
The electron-phonon coupling matrix elements $g_{{\bf q}\nu}({\bf k},i,j)$ are defined as \cite{wierzbowska,Heid2010,gustino-rmp}
\begin{equation}\begin{split}
g_{{\bf q}\nu}({\bf k},i,j)& =\\
=\sum_s &\sqrt{{\hbar\over 2M_s\omega_{{\bf q}\nu}}}
\langle\psi_{i,{\bf k+q}}| {dV_{\rm SCF}\over d {\hat u}_{\nu s} }\cdot
                   \hat \epsilon_{\nu s}|\psi_{j,{\bf k}}\rangle,
\label{eq:el-ph-matrix}
\end{split}\end{equation}
where $s$ enumerates atoms in the unit cell, $i,j$ are band indexes, $M_s$ is the atomic mass, $ {dV_{\rm SCF}\over d {\hat u}_{\nu s} }$ is a change of electronic potential calculated in self-consistent cycle due to the displacement ${\hat u}_{\nu s}$ of an atom $s$, $\hat{\epsilon}_{\nu s}$ is the $\nu$-th phonon polarization vector and $\psi_{i,{\bf k}}$ is the electronic wave function.
In eq.(\ref{eq:el-ph-matrix}) phonon frequency appears under the square root in the denominator.
The square of matrix elements are integrated over the Fermi surface and multiplied by frequency 
to contribute to the phonon linewidths $\gamma_{{\bf q}\nu}$:  \cite{wierzbowska,grimvall,gustino-rmp}:
\begin{equation}\label{eq:gamma}
\begin{split}
\gamma_{{\bf q}\nu} =& 2\pi\omega_{{\bf q}\nu} \sum_{ij}
                \int {d^3k\over \Omega_{\rm BZ}}  |g_{{\bf q}\nu}({\bf k},i,j)|^2 \\
                    &\times\delta(E_{{\bf k},i} - E_F)  \delta(E_{{\bf k+q},j} - E_F).
\end{split}
\end{equation}
This multiplication cancels the direct frequency dependence, thus the linewidths do not directly depend on $\omega_{{\bf q}\nu}$, rather than on the wavefunction matrix elements, which enter eq.(\ref{eq:el-ph-matrix}).

The summation of the phonon linewidths divided by frequency gives the Eliashberg function $\alpha^2F(\omega)$:
\begin{equation}\label{eq_a2F}
\alpha^2F(\omega) = {1\over 2\pi N(E_F)}\sum_{{\bf q}\nu} 
                    \delta(\omega-\omega_{{\bf q}\nu})
                    {\gamma_{{\bf q}\nu}\over\hbar\omega_{{\bf q}\nu}},
\end{equation}
which is then inversely proportional to $\omega_{{\bf q}\nu}$. 
Finally, the EPC constant $\lambda$ is calculated as the integral of the Eliashberg function, again divided by frequency \cite{grimvall},
\begin{equation}\label{eq_lambda_a2F}
\lambda=2\int_0^{\omega_{\rm max}} \frac{\alpha^2F(\omega)}{\omega} \text{d}\omega,
\end{equation}
which results in the proportionality of $\lambda \propto \frac{\gamma_{{\bf q}\nu}}{\omega_{{\bf q}\nu}^{2}}$, thus for majority of the electron-phonon superconductors pressure-induced lattice stiffening (increase in $\omega_{{\bf q}\nu}$) dominates, reducing $\lambda$.

\begin{table*}[t]
\caption{(Top part) Phonon modes' contributions $\lambda_{\nu}$ to the electron-phonon coupling constant $\lambda = \sum_{\nu} \lambda_{\nu}$ of $\mathrm{ThCoC_2}$ for 0 and 20 GPa. $\Delta \lambda$ is the difference between 20 GPa and 0 GPa results. (Bottom part) Mean phonon frequencies (in THz): global $\left<\omega\right>$ and for each of the phonon modes $\left<\omega\right>_{\nu}$. \label{tab_lambda_p0p20}}
\centering
\begin{ruledtabular}
\begin{tabular}{cccccccccccccc}
p (GPa) & $\lambda$ &$\lambda_{1}$ &$\lambda_{2}$ &$\lambda_{3}$ &$\lambda_{4}$ &$\lambda_{5}$ &$\lambda_{6}$ &$\lambda_{7}$ &$\lambda_{8}$ &$\lambda_{9}$ &$\lambda_{10}$ &$\lambda_{11}$ &$\lambda_{12}$ \\
\hline
0 & 0.5835 &0.0755 &0.0396 &0.0420 &0.0856 &0.0658 &0.0398 &0.0814 &0.1020 &0.0104 &0.0166 &0.0184 &0.0064 \\
20 & 0.6521 &0.1391 &0.0412 &0.0457 &0.0974 &0.0607 &0.0395 &0.1017 &0.0806 &0.0079 &0.0152 &0.0167 &0.0064 \\
$\Delta\lambda$ & 0.0686&	0.0636&	0.0016&	0.0037&	0.0118&	-0.0051&	-0.0003&	0.0203&	-0.0214&	-0.0025&	-0.0014&	-0.0017&	0.00\\
\hline
p (GPa) & $\left<\omega\right>$ & $\left<\omega\right>_{1}$ & $\left<\omega\right>_{2}$ & $\left<\omega\right>_{3}$ & $\left<\omega\right>_{4}$ & $\left<\omega\right>_{5}$ & $\left<\omega\right>_{6}$ & $\left<\omega\right>_{7}$ & $\left<\omega\right>_{8}$ & $\left<\omega\right>_{9}$ & $\left<\omega\right>_{10}$ & $\left<\omega\right>_{11}$ & $\left<\omega\right>_{12}$\\
0 & 10.6188 & 2.3185 & 2.6588 & 3.3310 & 4.6398 & 5.4076 & 7.2053 & 8.9207 & 10.1717 & 14.5968 & 16.0684 & 17.3748 & 34.7618\\
20 & 11.7611 & 2.4635 & 2.9604 & 3.8626 & 5.0101 & 6.1745 & 8.3831 & 10.1379 & 10.6632 & 16.1909 & 18.7990 & 20.1257 & 36.4349\\
\end{tabular}
\end{ruledtabular}
\end{table*}

To have a convenient single-number parameter which shows how the electronic contribution to the electron-phonon coupling changes with pressure, cutting off the direct dependence on the phonon frequency, one might calculate the sum of all phonon linewidths. 
The more straightforward way to do it is to calculate the integral of the Eliashberg function multiplied by frequency:
\begin{equation}
I=\int_0^{\omega_{\rm max}} \omega\cdot\alpha^2F(\omega){\rm d}\omega,
\label{eq:I}
\end{equation}
as it carries similar physical information.
Substituting eq.(\ref{eq:el-ph-matrix}-\ref{eq_a2F}) to the above formula, one can see that $I$ does not explicitly depend on phonon frequency~\cite{srm2}:
\begin{equation}
\begin{split}
I=&\frac{1}{2\pi\hbar N(E_F)} \int_0^{\omega_{max}} d\omega
                    \sum_{{\bf q}\nu}
                    \delta(\omega-\omega_{{\bf q}\nu}) \gamma_{\textbf{q}\nu}\\
=& \frac{1}{N(E_F)} \int_0^{\omega_{max}} d\omega
                    \sum_{{\bf q}\nu}
                    \delta(\omega-\omega_{{\bf q}\nu}) \\
&\times\sum_{ij}\int \frac{d^3k}{\Omega_{BZ}}\left|\sum_s \frac{1}{\sqrt{2M_s}} \langle\psi_{i,{\bf k+q}}| {dV_{\rm SCF}\over d {\hat u}_{\nu s} }\cdot
                   \hat \epsilon_{\nu s}|\psi_{j,{\bf k}}\rangle\right|^2 \\
&\times\delta(E_{{\bf k},i} - E_F)  \delta(E_{{\bf k+q},j} - E_F).\\
\end{split}
\end{equation}

\begin{figure*}[t]
	\centering
	\includegraphics[width=0.85\textwidth]{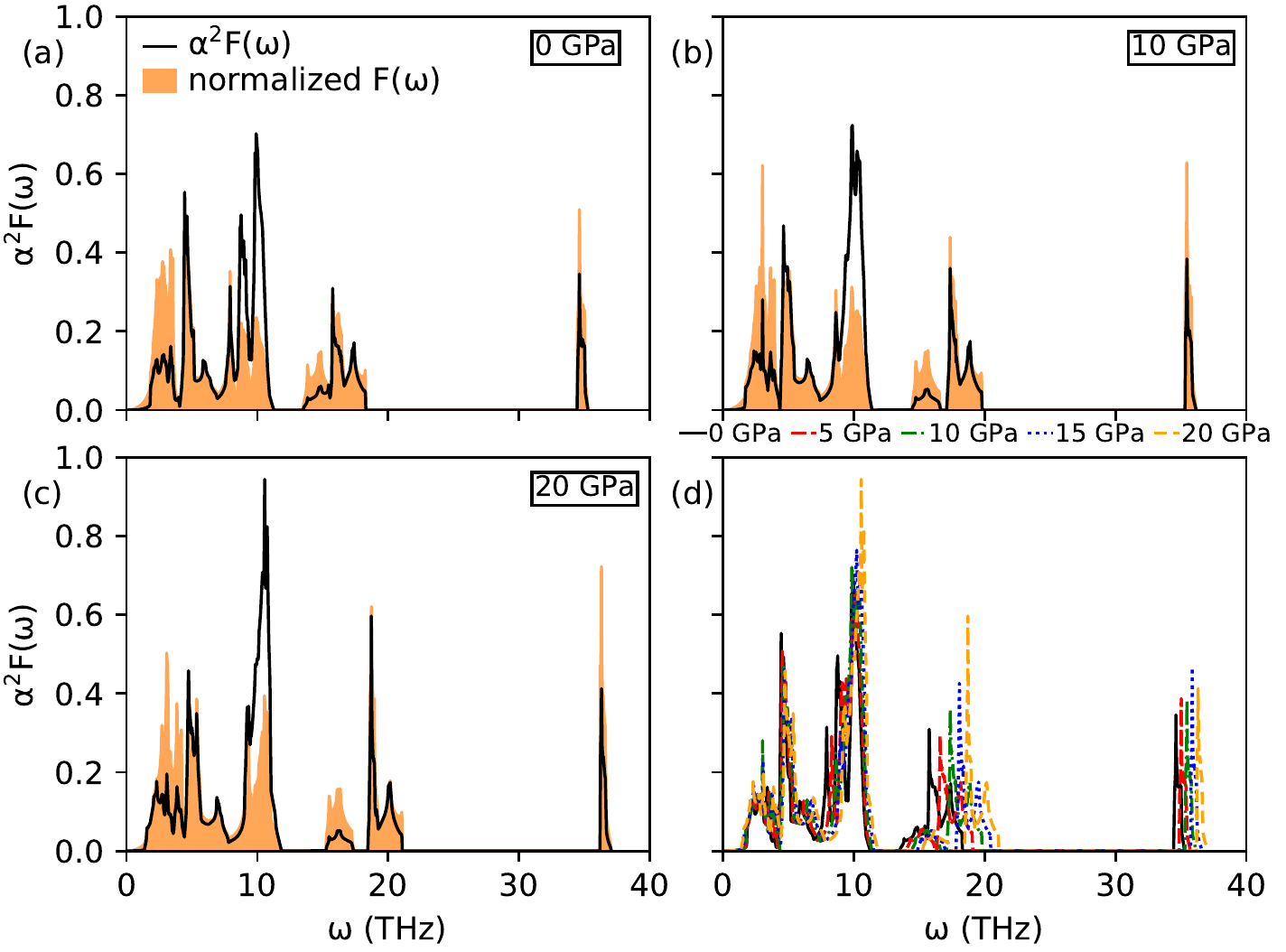}
	\caption{{Panels (a-c): Eliashberg function of $\mathrm{ThCoC_2}$ under pressures of 0, 10 and 20 GPa. Orange shading is the phonon density of states renormalized to have the same area as the corresponding Eliashberg function. In panel (d) Eliashberg functions for all pressures are gathered to visualize their relative evolution with pressure.}
	\label{plot_a2F_px2}}
\end{figure*}

In Fig.~\ref{plot_phbands} we see that some of the phonon linewidths are increasing with pressure, especially for the softened acoustic part of the spectrum between T, Z, and A points, which shows that the softening is associated with the enhanced electron-phonon coupling.
On the other hand, for the optic Co and C modes, e.g., in the X-$\Gamma$ direction, $\gamma_{\pmb{q}\nu}$ are decreasing. 
Eliashberg functions for selected pressures are presented in Fig.~\ref{plot_a2F_px2} and an increase in the lower-frequency part can be seen.
The overall effect on the electron-phonon coupling is the resultant of competing factors: the increase of the ''electronic'' $I$ integral, plotted in Fig.~\ref{plot_meanfreqs}(c), and increase in the average phonon frequency, shown in Fig.~\ref{plot_meanfreqs}(a). 
The tendency towards stronger coupling takes over and the EPC constant $\lambda$ is increasing with pressure with a rate of about $3.5\times 10^{-3}$/GPa, as shown in  Fig.~\ref{plot_meanfreqs}(d). This is 20\% slower than found for LaNiC$_2$, where it was equal to $4.4\times 10^{-3}$/GPa~\cite{lanic2}.

\begin{figure}[t]
	\centering
	\includegraphics[width=0.80\columnwidth]{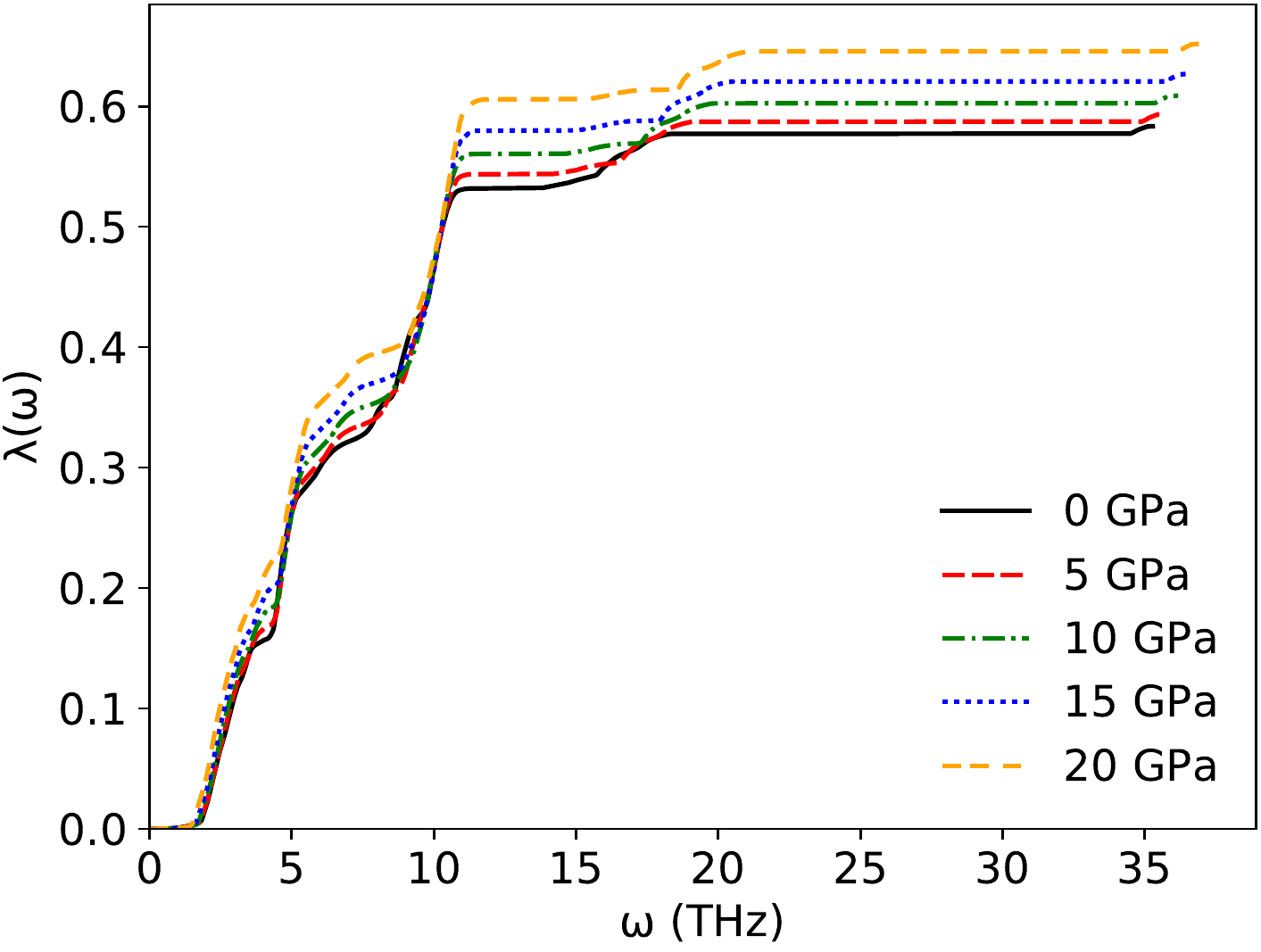}
	\caption{Frequency distribution of the electron-phonon coupling constant of $\mathrm{ThCoC_2}$ {calculated as $\lambda(\omega)=2\int_0^{\omega} \frac{\alpha^2F(\omega')}{\omega'} \text{d}\omega',$ }. \label{plot_clamb}}
\end{figure}

Analyzing the increase in $\lambda$ in more details, Fig. ~\ref{plot_phmod_a2F_p0p20} additionally compares the p = 0 and 20 GPa Eliashberg functions decoupled over all the phonon modes [panels (a,b)], as well as $\alpha^2F(\omega)/\omega$ in panel (c). 
The substantial increase of the Eliashberg function for the first acoustic mode is well visible. For other modes
generally the increase in phonon frequencies is accompanied by the increase in $\alpha^2F(\omega)$ values due to the larger phonon linewidths. This gives a larger $\alpha^2F(\omega)/\omega$ function, whose integral determines the value of $\lambda$ via eq.(\ref{eq_lambda_a2F}).
Electron-phonon coupling constants for each of the 12 phonon modes $\lambda_i$, computed based on the data presented in Fig.~\ref{plot_phmod_a2F_p0p20}(a,b) are collected in Table~\ref{tab_lambda_p0p20}. 
The pressure changes in the cumulative electron-phonon coupling constant $\lambda(\omega)$ are shown in Fig. \ref{plot_clamb}.
What we can conclude is that the increase in $\lambda$ is mainly driven by the increase in $\lambda_1$ which is contributed by the lowest acoustic phonon mode, 
the one which softens under pressure in some regions of the Brillouin zone. 
As a consequence, the average frequency of this mode (see $\langle \omega \rangle_1$ in Table~\ref{tab_lambda_p0p20}) increases only a little, allowing for the increase in $\lambda_1$.
Changes in $\lambda_{\nu}$ of the other modes generally compensate each other when summed over, thus the change in $\lambda_1$ is a key factor for the $\lambda$ enhancement.
The increase in $\lambda$ will tend to increase the critical temperature $T_c$.

The second important parameter in terms of $T_c$ is the characteristic phonon frequency, which sets the energy window for the electron-phonon interaction, and to which $T_c$ is proportional to. That is the Debye frequency in the BCS theory and in the McMillan $T_c$ formula~\cite{mcmillan}. 
The pressure-induced increase in the average phonon frequency suggests that this parameter should increase as well.
However, in the more accurate Allen-Dynes formula:
\begin{equation}
	k_B T_c = \frac{\hbar \omega_{\rm ln}}{1.2} \exp \left(- \frac{1.04(1+\lambda)}{\lambda-\mu^*(1+0.62\lambda)} 
\right), \label{eq_Allen_Dynes}
\end{equation}
which is an analytic approximated solution of the isotropic Eliashberg gap equations~\cite{allen}, the characteristic frequency is  the ''logarithmic average'', 
defined on the basis of the Eliashberg function:

\begin{figure}[t]
	\centering
	\includegraphics[width=0.99\columnwidth]{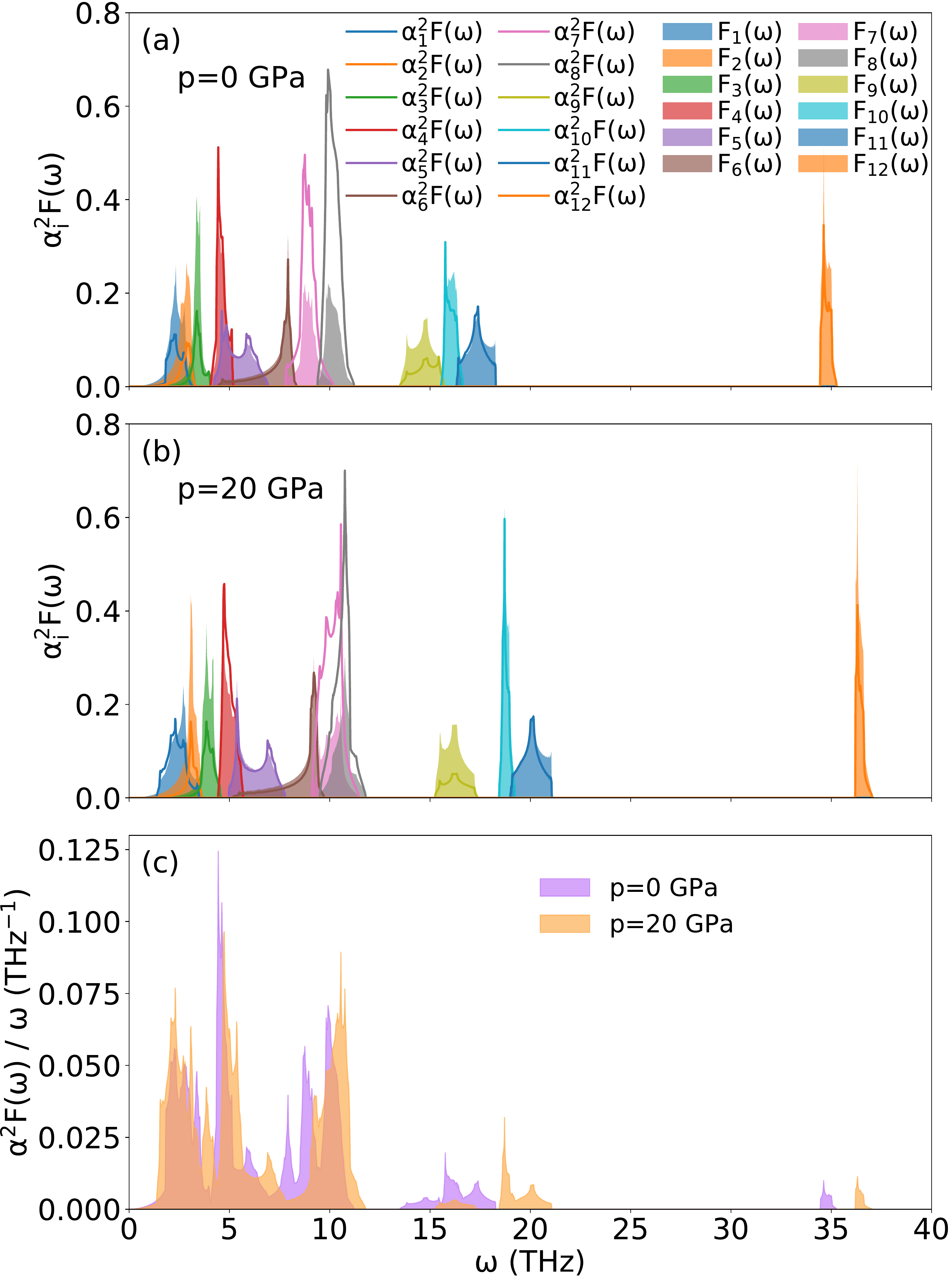}
	\caption{Partial Eliashberg functions and normalized partial phonon density of states of $\mathrm{ThCoC_2}$ for (a) 0 GPa and (b) 20 GPa. (c) Eliashberg functions devided by frequency for 0 and 20 GPa. \label{plot_phmod_a2F_p0p20}}
\end{figure}

\begin{equation}\label{eq:wln}
	\omega_{\rm ln} = \exp \left( \frac{2}{\lambda}\int_0^{\omega_{max}} \alpha^2F(\omega) \ln \omega \frac{d\omega}{\omega} \right).
\end{equation}

Increases in both the average phonon frequency and the EPC constant $\lambda$ results in a non-monotonic evolution of $\omega_{\rm ln}$, 
shown in Fig.~\ref{plot_meanfreqs}(b). 
Logarithmic average increases at low pressures, reaches maximum at 5~GPa and then decreases.

\subsection{Superconductivity}
\label{sec:eliashberg}

The results obtained in the preceding sections allow to simulate the pressure effects on superconductivity in ThCoC$_2$ for the electron-phonon coupling scenario. 
As we have discussed in Ref.~\cite{Kuderowicz2021}, the experimental value of the critical temperature $T_c$ at zero pressure can be obtained from the calculated Eliashberg function and the standard Allen-Dynes eq.(\ref{eq_Allen_Dynes}), 
if the Coulomb pseudopotential parameter is set to $\mu^* \simeq 0.16$. 
The need of using this higher than typical (0.10 - 0.13) value 
is most likely related to the more complex nature of the superconducting phase than covered by the Allen-Dynes formula, derived based on the isotropic Eliashberg formalism.
Similarly, by direct numerical solving the isotropic Eliashberg equations, we have reproduced~\cite{Kuderowicz2021} the experimental value of $T_c$ for $\mu^* = 0.268$ when $\alpha^2F(\omega)$ computed with SOC was used.\footnote{The value of $\mu^*$  used for solving the Eliashberg gap equations, if is to be compared to the parameter used with the Allen-Dynes formula, has to be re-scaled due to the dependence of $\mu^*$ on the numerical cutoff frequency: $\frac{1}{\widetilde{\mu^*}} = \frac{1}{\mu^*} + \ln\left(\frac{\omega_c}{\omega_{\rm max}}\right)$. That gives $\widetilde{\mu^*} = 0.195$, see Refs.~\cite{allen,Kuderowicz2021} for further details.}
This and the deviation of the calculated thermodynamic properties of the superconducting phase (temperature dependence of the electronic specific heat, magnetic field penetration depth, and critical field) pointed to the conclusion that the superconductivity in ThCoC$_2$ goes beyond the isotropic Eliashberg picture, nevertheless with a high probability of the electron-phonon coupling mechanism~\cite{Kuderowicz2021}.
As we are, however, able to reproduce the magnitude of $T_c$ at p = 0 accepting the enhanced $\mu^*$ values, 
we may simulate the pressure effect on $T_c$ by keeping the zero-pressure $\mu^*$ 
that reproduces the initial $T_c$.
 
\begin{figure}[t]
\includegraphics[scale=.55]{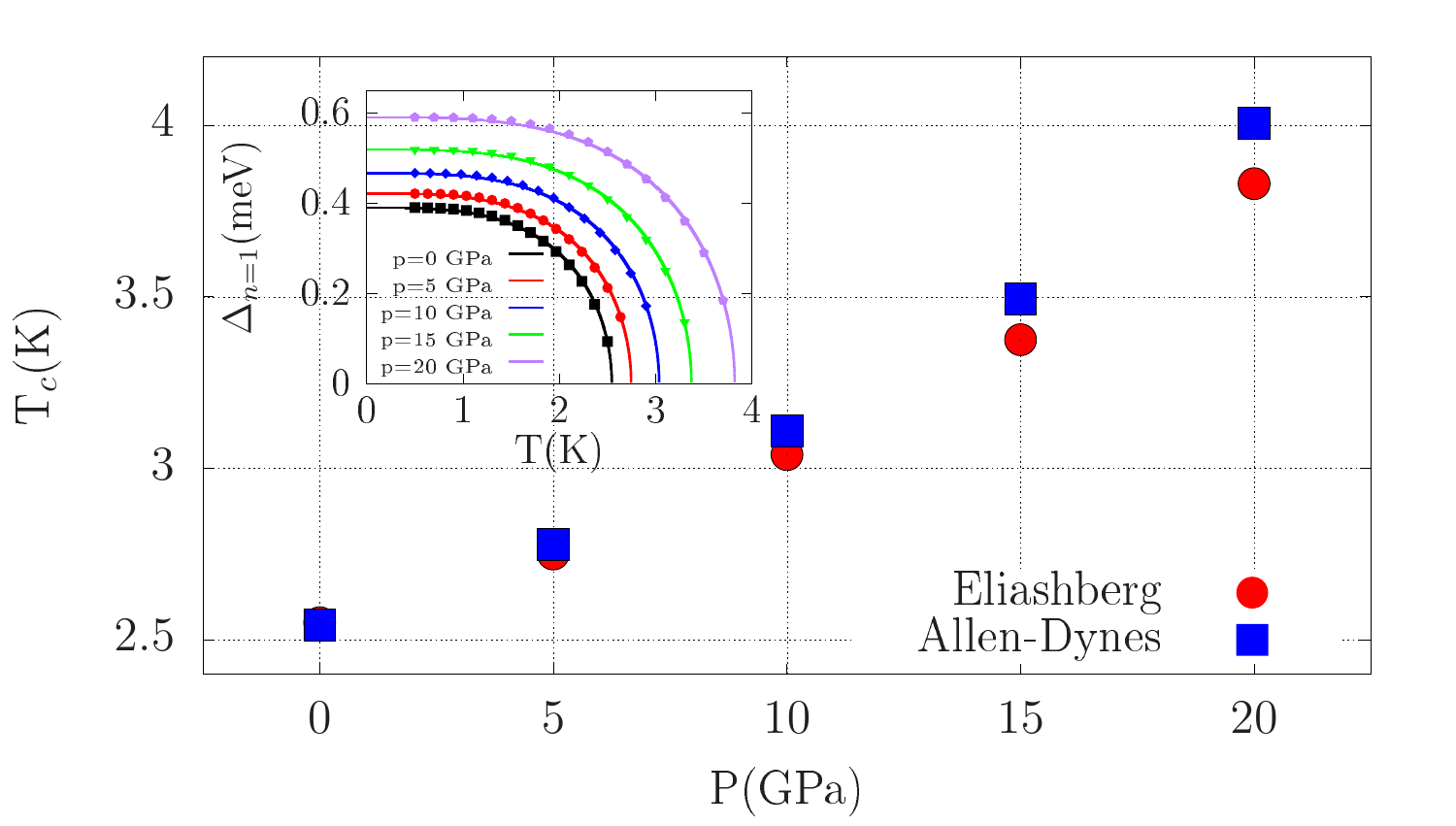}
\caption{Critical temperature $T_c$ as a function of the pressure p, determined from the Allen-Dynes formula (blue squares) and from the numerical solving of Eliashberg equations (red dots). In both cases $\mu^*$ was adjusted to reproduce the experimental zero-pressure $T_c$ value.
Inset: Temperature dependence of the superconducting order parameter $\Delta_{n=1}$ for different values of p.}
\label{fig:Eliash1}
\end{figure}

Starting with $T_c$ computed using the Allen-Dynes formula,  
$T_c$ increases almost linearly with pressure, as shown in Fig.~\ref{fig:Eliash1} (blue squares).  $\mu^* = 0.161$ is used here, as it reproduces the experimental $T_c = 2.55$~K at 0 GPa when SOC is neglected.
In spite of the decline of the density of state $N(E_F)$, in general detrimental to superconductivity, we observe increase of the critical temperature with a pressure, caused by the enhancement of EPC.
The increase in $\lambda$ takes over the decrease in $\omega_{\rm ln}$ observed above 5 GPa, thus the critical temperature increases monotonically with an approximately constant ratio of 0.073~K/GPa, reaching 3.11~K at 10~GPa and 4.01~K at 20~GPa.
That is quite a strong pressure effect, experimentally accessible for verification even under moderate pressures of several GPa. 
Following the trend in $\lambda(p)$, the magnitude of the pressure-induced change in $T_c$ in ThCoC$_2$ is lower comparing to that computed for  LaNiC$_2$~\cite{lanic2}, where the ratio was 0.129~K/GPa.

Further on, we simulate how the $T_c$ and other thermodynamic properties would change with pressure when computed using the Eliashberg isotropic gap equations. 
That will allow to conclude on the deviations of superconductivity in ThCoC$_2$ under pressure from the isotropic state once the experimental works are reported.
A full description of the applied theoretical model {with mathematical formulas} can be found in Ref.~\cite{Kuderowicz2021}.
Here we keep the same numerical parameters used in solving the Eliashberg equations, i.e., the cut-off frequency $\omega _c = 4 \omega _{max}$~\cite{Carbotte1990} and the number of Matsubara frequencies $M=6500$. 
As mentioned, the Coulomb pseudopotential has been chosen in a way to reproduce the critical temperature $T_c=2.55$~K for p = 0. In the model without the spin-orbit coupling, this is $\mu^*=0.257$, which corresponds to re-scaled Allen-Dynes formula value ~\cite{allen, Kuderowicz2021} $\widetilde{\mu^*} = 0.189$, still above the commonly used approximations of $0.10-0.13$~\cite{Morel1962}.
The changes of the critical temperature under pressure are presented in Fig.~\ref{fig:Eliash1} (red dots), and are 
slightly smaller, comparing to predictions from the Allen-Dynes formula (rate of 0.64~K/GPa), nevertheless the substantial increase in $T_c$ is predicted as well.

The temperature dependence of $\Delta_{n=1}(T)$ for different pressures are presented in the inset of Fig.~\ref{fig:Eliash1} and undergo the following formula
\begin{equation}
 \Delta(T)=\Delta(0)\sqrt{1-\left ( \frac{T}{T_c} \right )^\Gamma},
 \label{eq:delT}
\end{equation}
with $\Gamma$ which only slightly depends on  p, $\Gamma=3.28\pm 0.01$, remaining  close to that predicted from the BCS theory, $\Gamma_{BCS} \approx 3.0$. 
When increasing the pressure, the dimensionless ratio $R_{\Delta} = 2 \Delta (0) / k_B T_c$ changes in the range $3.55-3.58$, being close to the BCS value of $3.53$.

\begin{figure}[t]
\includegraphics[scale=.55]{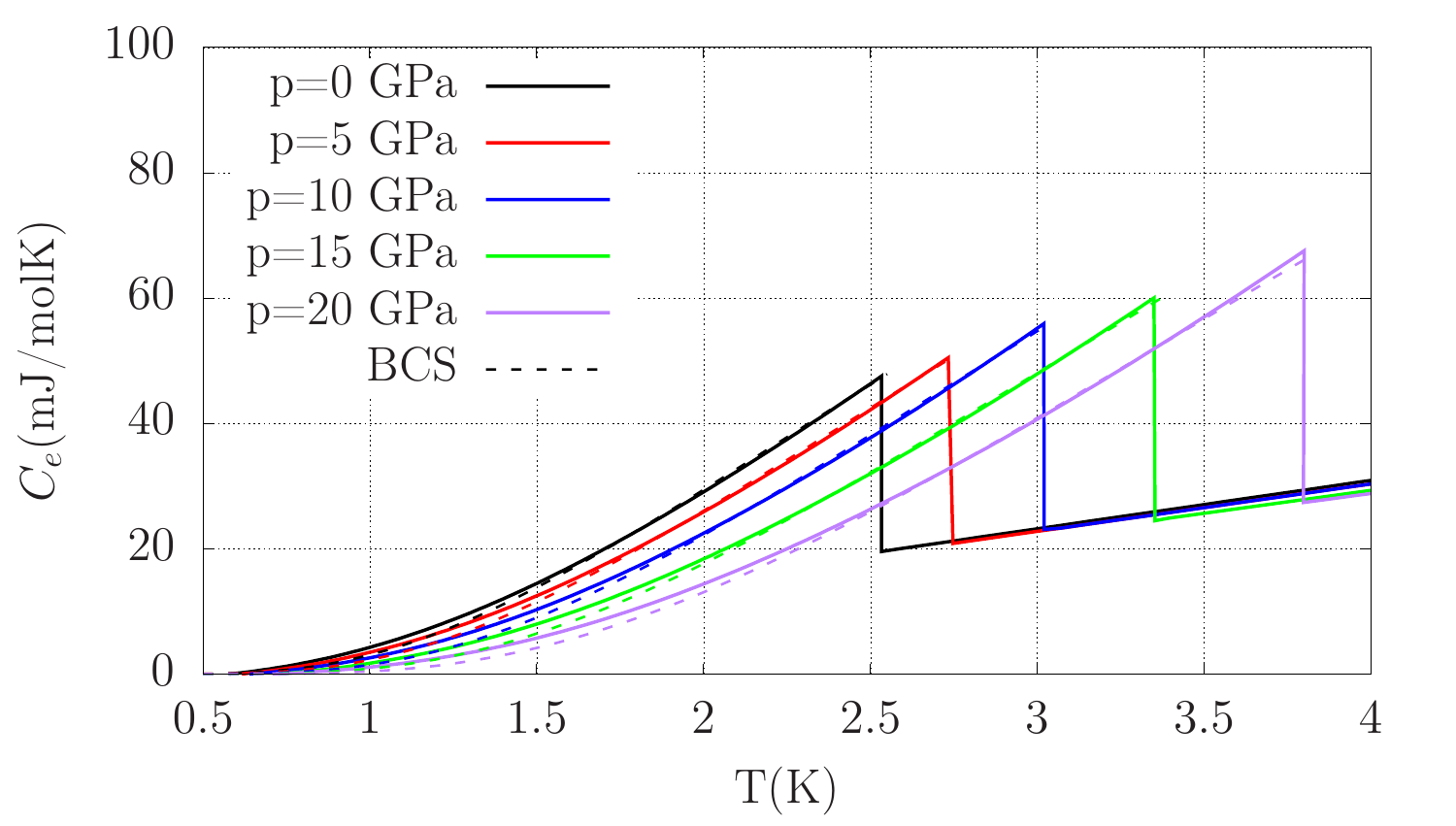}
\caption{Temperature dependence of the electronic specific heat $C_e(T)$ for different pressure values. Dashed lines represent the BCS predictions. Results for $\mu^*=0.257$.}
\label{fig:C}
\end{figure}

The self-consistent solution of the Eliashberg equations can then be used to calculate the electronic specific heat in the superconducting state {$C_e^S$ \cite{Kuderowicz2021}. Above $T_c$, the specific heat of the normal state is given by 
$C_e^N(T) = \frac{\pi^2}{3}k_B^2 N(E_F)(1+\lambda)T$.}
The temperature dependences of the specific heat for different pressure values are presented in Fig.~\ref{fig:C}. 
For comparison, by dashed lines, we have also marked the BCS curves, which predict the exponential behavior of the specific heat at low temperatures. In this regime, as shown in Fig.~\ref{fig:C}, the Eliashberg solutions slightly deviate from the BCS curves and do not exhibit the $C_e \propto \exp[-\Delta(0)/k_BT]$ dependence. At high temperatures, $C_e^S(T)$ is close to BCS relation, reaching the jump of reduced specific heat at $T_c$, $\Delta C_e/\gamma T_c$ from $1.37$ for p = 0 to $1.47$ for $p=20$ GPa, almost equal to the weak-coupling BCS limit of $1.43$. 

\begin{figure}[t]
\includegraphics[scale=.55]{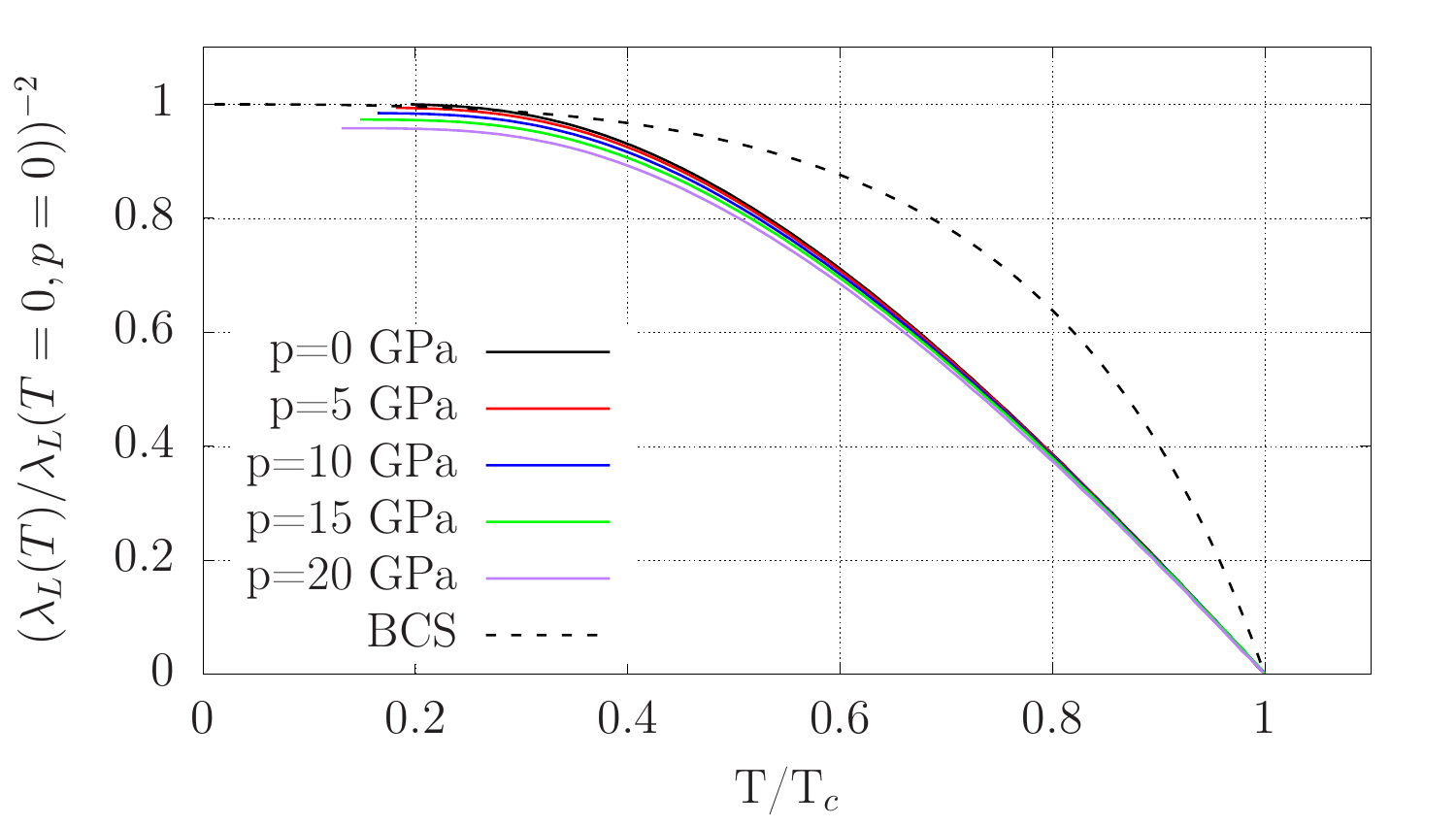}
\caption{ The normalized London penetration depth $\lambda _L/\lambda_L(T=0,p=0)$ as a function of the normalized temperature $T/T_c$.}
\label{fig:ld}
\end{figure}

Although electronic specific heat exhibits a temperature dependence close to ordinary BCS theory, the London penetration depth $\lambda _L$ behaves differently. In Fig.~\ref{fig:ld} we present $\lambda _L/\lambda_L(T=0,p=0)$ calculated from {the Eliashberg formalism~\cite{Kuderowicz2021,Carbotte1990}.}
Regardless of pressure, $\lambda _L$ significantly deviates from the BCS prediction, which is an effect of the retardation processes included in the Eliashberg theory.
With increasing pressure, the London penetration depth decreases slightly in the low temperature regime. 

\begin{figure}[t]
\includegraphics[scale=.55]{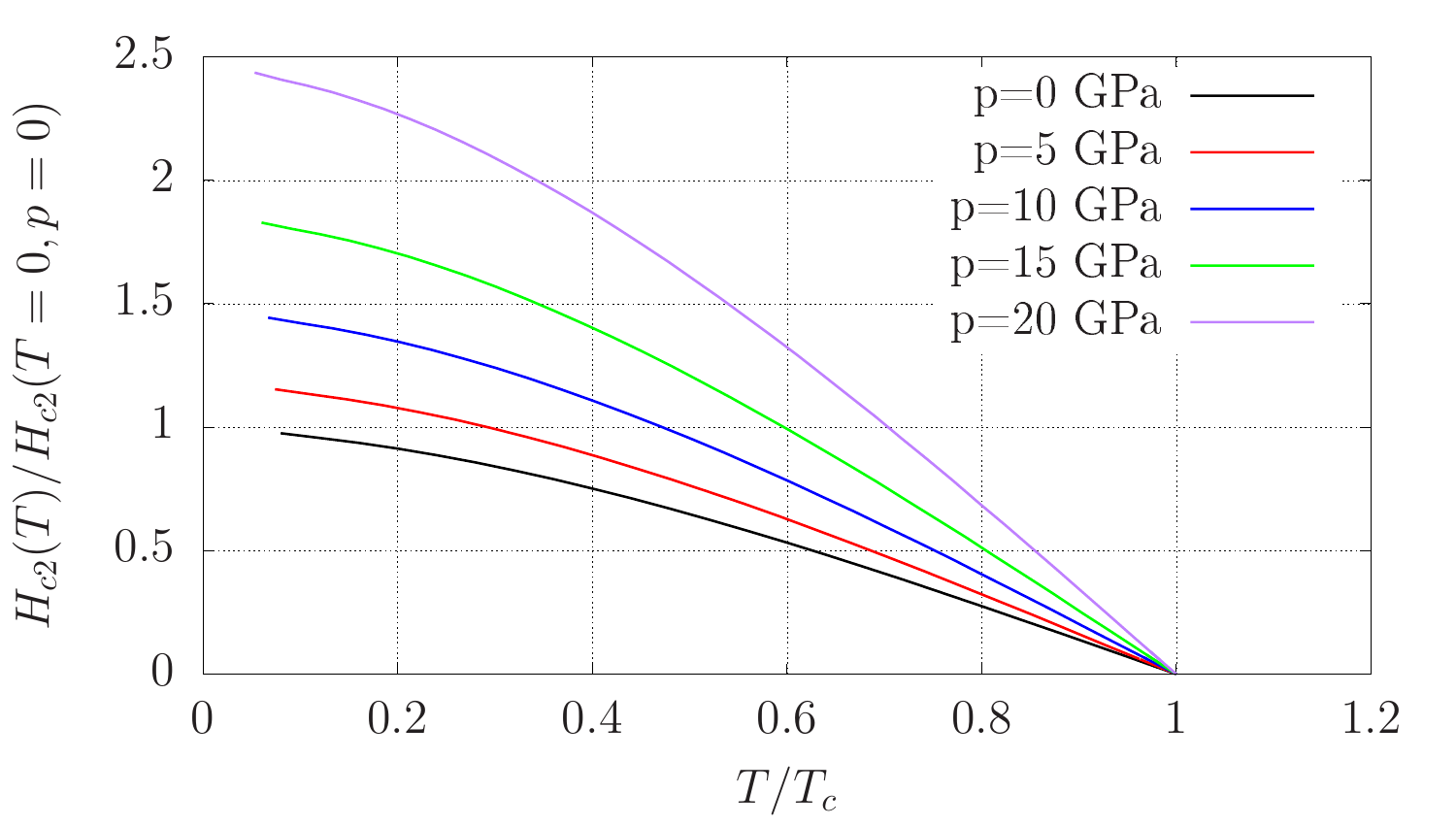}
\caption{ Normalized upper critical field $H_{c2}/H_{c2}(T=0,p=0)$ as a function of normalized temperature $T/T_c$.}
\label{fig:hc}
\end{figure}

{Finally, we have also analyzed how the upper critical field, $H_{c2}$,
changes within the pressure, if evaluated within the Eliashberg model~\cite{Kuderowicz2021,Carbotte1990}}.
Our calculations for zero pressure~\cite{Kuderowicz2021} showed that ThCoC$_2$ was close to the clean limit, thus in  present  calculations we use the clean limit, {assuming zero scattering rate.}

The obtained temperature dependence of the  upper critical field normalized to $H_{c2}$ at $T=0$ and p = 0 is shown in Fig.~\ref{fig:hc}. In opposite to the penetration depth which does not change significantly with the pressure, the upper critical field increases more than twice  when the pressure changes from $0$ to $20$~GPa. Similarly as in the case of the critical temperature, it results from the increase of the EPC induced by pressure.

\section{Summary}
\label{sec:summary}
We have analyzed the pressure evolution of the electronic structure, phonons and the electron-phonon coupling in ThCoC$_2$ in the range p = 0 - 20 GPa.
Calculations showed that the density of states at the Fermi level slowly decreases with p. The band splitting induced by the spin-orbit coupling increases with p, reaching the value of 185~meV at 20 GPa. This suggests that under elevated pressure the non-BCS features in the thermodynamic properties of the superconducting phase in ThCoC$_2$ may be even more pronounced than at ambient pressure.
The electron-phonon coupling is enhanced under pressure due to the increase in the phonon linewidths.
The overall EPC parameter $\lambda$ increases  from 0.583 at 0~GPa to 0.652 at 20~GPa, which is strongly related to the softening of the lowest acoustic phonon mode.
If the superconducting pairing in ThCoC$_2$ is based on the electron-phonon coupling, this should raise the superconducting critical temperature and the upper magnetic critical field. 
When the isotropic Eliashberg formalism is used to calculate $T_c$, enhancement of the electron-phonon coupling strength
results in a significant increase of $T_c$, with a rate of 0.064 - 0.073 K/GPa. 

\section*{Acknowledgements}
This work was supported by the National Science Centre (Poland), project no. 2017/26/E/ST3/00119 and in part by the PL-Grid infrastructure.

\bibliography{refs}

\vspace*{1054pt}

\section{Supplemental Material}

\begin{figure*}[htb]
	\centering
	\includegraphics[width=0.50\textwidth]{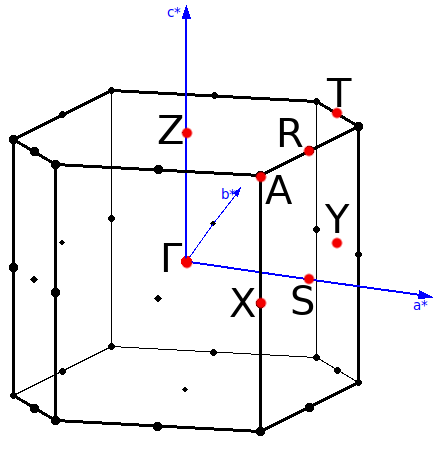}
	\caption{High symmetry points in the first Brillouin Zone of a base centered orthorhombic cell. \label{plot_BZ}}
\end{figure*}

\begin{figure*}[htb]
	\centering
	\includegraphics[width=0.89\textwidth]{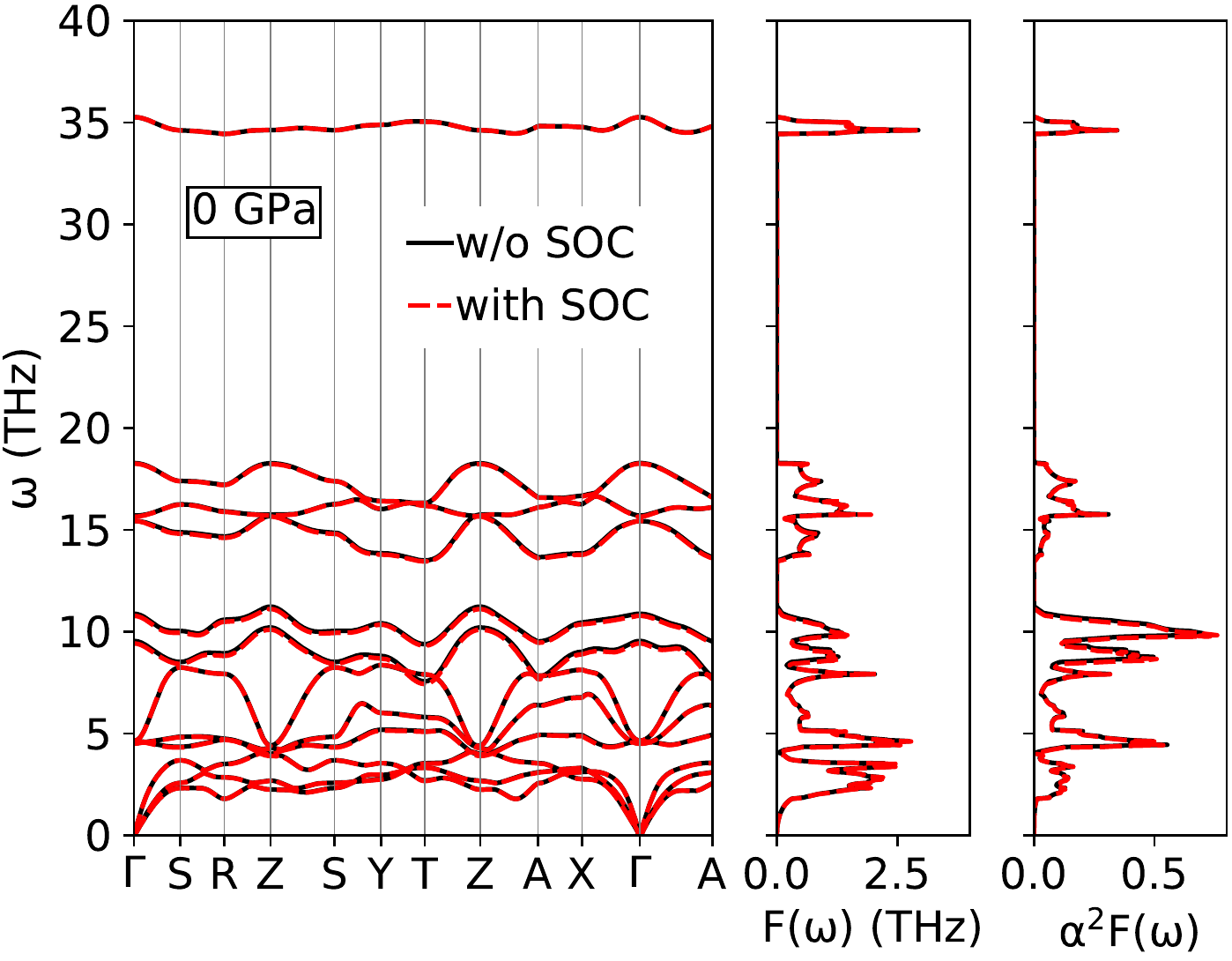}
	\caption{The effect of SOC on phonon dispersion relation, density of states and Eliashberg function of $\mathrm{ThCoC_2}$ under 0 GPa. \label{plot_php0_so}}
\end{figure*}

\begin{figure*}[htb]
	\centering
	\includegraphics[width=0.89\textwidth]{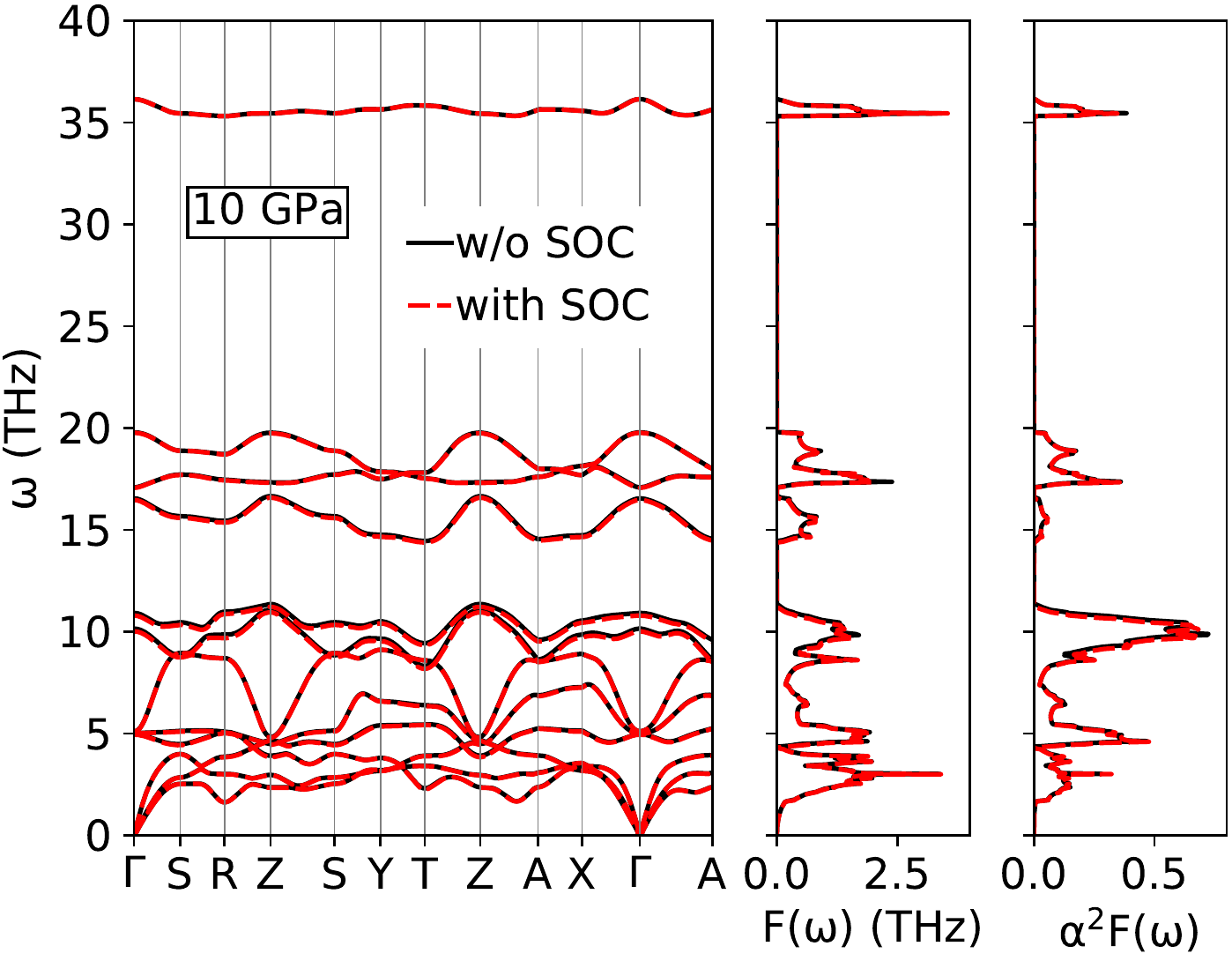}
	\caption{The effect of SOC on phonon dispersion relation, density of states and Eliashberg function of $\mathrm{ThCoC_2}$ under 10 GPa. \label{plot_php10_so}}
\end{figure*}

\begin{figure*}[htb]
	\centering
	\includegraphics[width=0.99\textwidth]{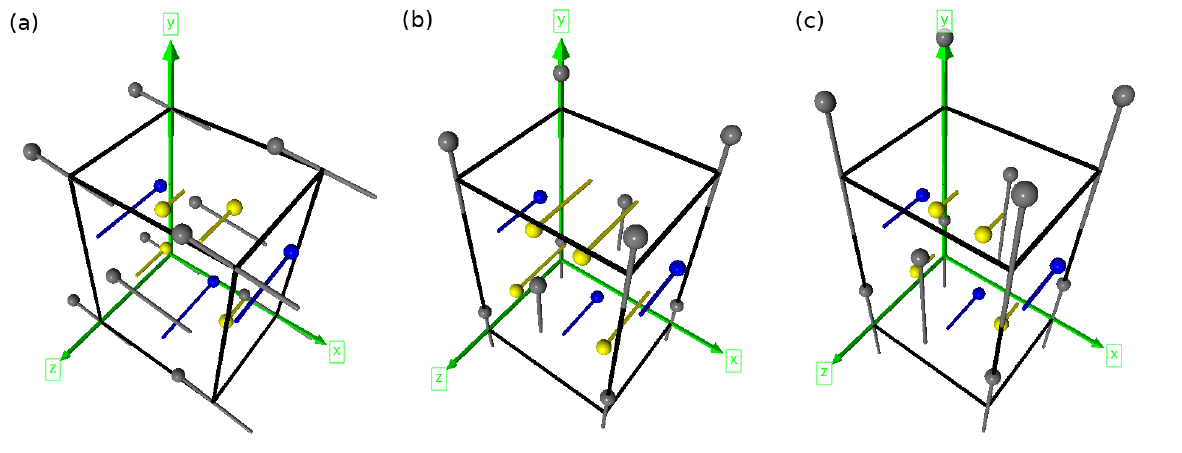}
	\caption{Visualization of $\mathrm{ThCoC_2}$ atoms' vibrations in the first acoustic mode at (a) R, (b) T and (c) Z $\mathbf{q}$-points. Gray balls are thorium atoms, blue are cobalt and yellow are carbon atoms. \label{plot_disp_RT}}
\end{figure*}

\begin{figure*}[htb]
	\centering
	\includegraphics[height=0.99\textheight]{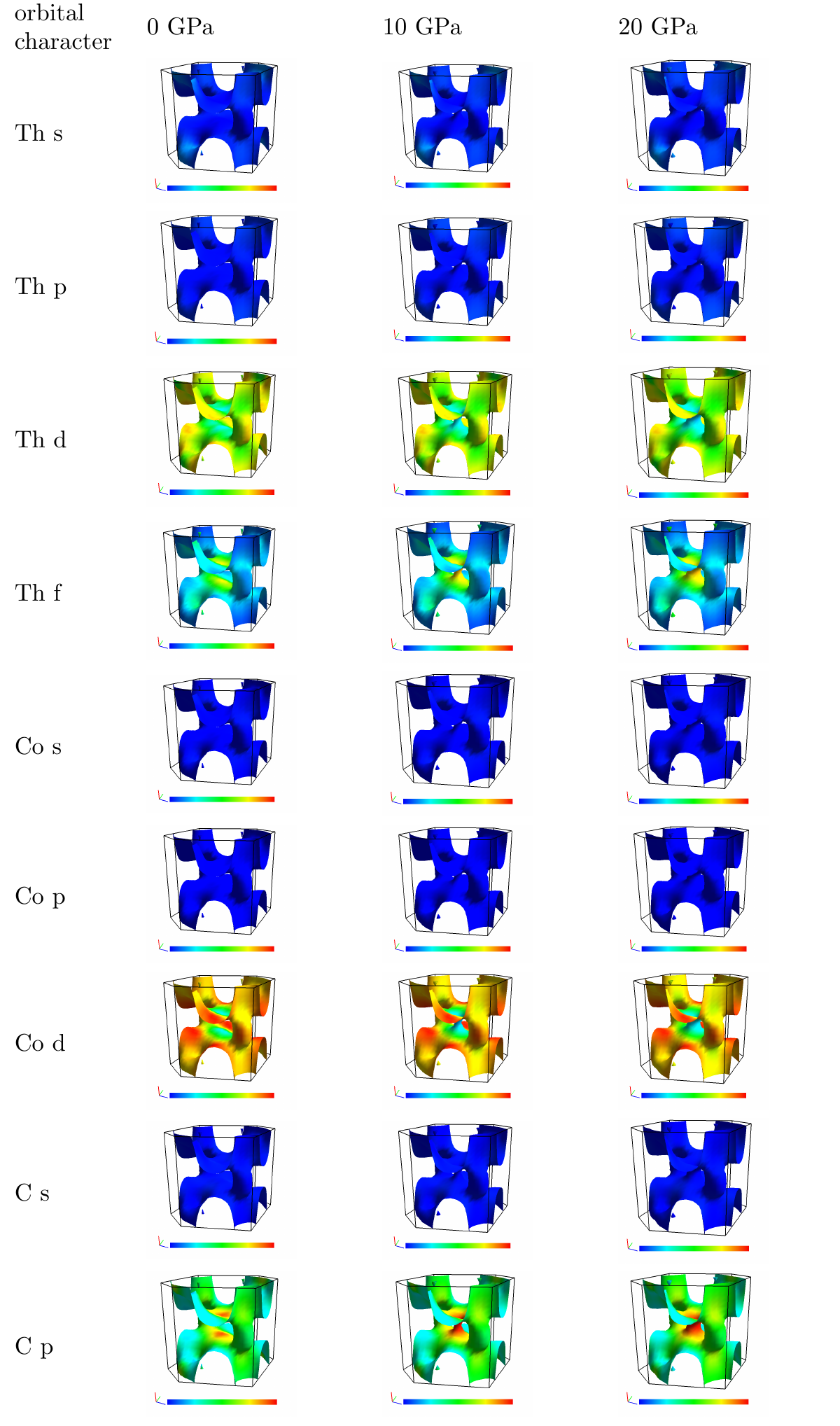}
	\caption{Orbital character of states forming the Fermi surface of $\mathrm{ThCoC_2}$. Color scale is from 0.0 (blue) to 0.45 (red), and all orbital contributions at each $k-$point sum to 1.0. \label{plot_fs_orbchar}}
\end{figure*}

\noindent
Supplemental Material contains:\\ \\
Fig. \ref{plot_BZ} of the Brillouin zone with the high symmetry points.\\ 
Fig. \ref{plot_php0_so} which compares phonon dispersion relations, phonon density of states and Eliashberg function of ThCoC$_2$ computed for p = 0 GPa with and without spin-orbit coupling (SOC). \\
Fig. \ref{plot_php10_so} which compares phonon dispersion relations, phonon density of states and Eliashberg function of ThCoC$_2$ computed for p = 10 GPa with and without spin-orbit coupling (SOC). \\
Fig. \ref{plot_disp_RT} with the displacement patterns of atoms vibrating in the first acoustic phonon mode at R, T anz Z $\mathbf{q}$ points at p = 0 GPa. Amplitude of vibrations is exaggerated. Atomic displacements remain similar for larger pressures. \\
Fig. \ref{plot_fs_orbchar}

Figures \ref{plot_php0_so} and \ref{plot_php10_so} show that the spin-orbit coupling have a minor effect on the phonon dispersion relations and Eliashberg functions of ThCoC$_2$ also under elevated pressure.

Orbital characters of the main Fermi surface sheet for pressures of 0, 10  and 20 GPa are shown in Fig. \ref{plot_fs_orbchar}. Contribution to Fermi surface is mostly from Th-6d, Co-3d and C-2p orbitals, with some contribution from Th-f states, in agreement with contributions to total DOS at $E_F$ discussed in the main text. 
The anisotropy of orbital character of FS is moderate and is enhanced under pressure -- the tube which is formed in the central part of FS above 10 GPa has mostly C-p and Th-f character, which were more uniformly distributed at ambient pressure. Also the relative contribution to FS from 5f states of Th is growing under pressure. That may enhance anisotropy of the superconducting properties of ThCoC$_2$.

\end{document}